\documentclass[11pt]{article}
\usepackage{verbatim,amsmath,amssymb}
\usepackage{epsfig,float,color}
\usepackage{epstopdf}
\usepackage{geometry}
\usepackage{setspace}
\usepackage{wrapfig}
\usepackage{hyperref}
\usepackage[utf8]{inputenc}
\usepackage{graphicx}
\usepackage{flushend}
\usepackage{tikz,pgfplots}
\usepackage[font={footnotesize}]{caption}
\usepackage[font={footnotesize}]{subcaption}
\usepackage{footnote}
\usepackage{multicol}
\usepackage{multirow}
\usepackage{enumitem}   
\usepackage{algorithm}%
\usepackage{algorithmicx}%
\usepackage{algpseudocode}%
\usepackage{soul}
\usepackage{framed}
\usepackage[titletoc,title]{appendix}
\usepackage{bm}
\usepackage{siunitx}
\usepackage[makeroom]{cancel} 
\usepackage{hyperref}
\usepackage{bigints} 
\newcommand{\mb}[1]{\mathbf{#1}} 
\newenvironment{rcases}
{\left.\begin{aligned}}
	{\end{aligned}\right\rbrace}

\definecolor{darkblue}{rgb}{0,0,1}
\hypersetup{pdftex=true, colorlinks=true, breaklinks=true, linkcolor=blue, menucolor=blue, citecolor=blue, urlcolor=blue}

\definecolor{beaublue}{rgb}{0.94, 0.97, 1.0}
\newcommand{\trr}[1]{{#1}^{\!\top}}
\newcommand{\pd}[2]{\frac{\partial #1}{\partial #2}}
\newcommand{\inv}[1]{{#1}^{\text{-}1}}

\usepackage{tikz}
\usetikzlibrary{positioning, arrows.meta}
\tikzset{%
	myarrow/.style = {-Stealth, shorten >=5pt}
}

\usepackage{listings}
\definecolor{LightCyan}{rgb}{0.88,1,1}
\definecolor{mygreen}{RGB}{28,172,0} 
\definecolor{mylilas}{RGB}{170,55,241}
\begin{document}
	
	\begin{center}
		\Large{\bf{Normalized field product approach: A parameter-free density evaluation method for close-to-binary solutions in topology optimization with embedded length scale}}\\
		
	\end{center}
	
	\begin{center}
		
			\large{Nikhil Singh\,$^{\dagger}$, Prabhat Kumar$\,^{\star}\,$\footnote[1]{Corresponding author: \url{pkumar@mae.iith.ac.in}} and Anupam Saxena\,$^{\dagger,\ddagger}$} \\
		\vspace{4mm}
		
		 \small{\textit{$\dagger$Department of Mechanical Engineering, Indian Institute of Technology Kanpur, UP 208016, India}}\\
		 
		\small{\textit{$\star$Department of Mechanical and Aerospace Engineering, Indian Institute of Technology Hyderabad, Telangana
				 502285, India}}\\
		
        \small{\textit{$\ddagger$IGMR, RWTH University, Aachen, Germany}}\\
        
		\vspace{4mm}

	\end{center}
	
	\vspace{1mm}
	\rule{\linewidth}{.15mm}
	{\bf Abstract:}
	This paper provides a normalized field product approach for topology optimization to achieve close-to-binary optimal designs. The method employs a parameter-free density measure that implicitly enforces a minimum length scale on the solid phase, allowing for smooth and transition-free topologies. The density evaluation does not rely on weight functions; however, the related density functions must have values between 0 and 1. The method combines the SIMP scheme and the introduced density function for material stiffness interpolation. The success and efficacy of the approach are demonstrated for designing both two- and three-dimensional designs, encompassing stiff structures and compliant mechanisms. The structure's compliance is minimized for the former, while the latter involves optimizing a multi-criteria objective. Numerical examples consider different volume fractions, length scales, and density functions. A volume-preserving smoothing and resolution scheme is implemented to achieve serrated-free boundaries. The proposed method is also seamlessly  extended with advanced elements for solving 3D problems. The optimized designs obtained are close to binary without any user intervention while satisfying the desired feature size on the solid phase. \\
	
	{\textbf {Keywords:} {Topology Optimization; Compliant Mechanisms; Stiff strucutres;  Mesh independent solutions, Minimum length scale}

	\vspace{-4mm}
	\rule{\linewidth}{.15mm}
 \section{Introduction}
 These days, topology optimization (TO) has become one of the most used systematic approaches to solving a wide variety of design optimization problems for different applications involving single and/or multi-physics concepts. The method furnishes optimum material layout within the given design domain $\Omega$ by extremizing the desired objective under a given set of constraints. In density-based methods~\cite{bendsoe2003topology}, presence or absence of material at a point is expressed using the \textit{density field}, $\rho(\boldsymbol{\mathrm{X}})$. $0 \leq \rho(\boldsymbol{\mathrm{X}}) \leq 1$ for $\boldsymbol{\mathrm{X}} \in \Omega$, where $\rho(\boldsymbol{\mathrm{X}}) = 1$ and $\rho(\boldsymbol{\mathrm{X}}) = 0$ imply presence and absence of material at point (finite element) $\boldsymbol{\mathrm{X}}$, respectively~\cite{bendsoe2003topology}. In addition to density-bases methods, several other approaches, e.g., level-set methods~\cite{allaire2004structural}, evolutionary structural optimization~\cite{huang2010evolutionary}, featured-based methods~\cite{wein2020review,saxena2011topology}, TO of Binary Structures method~\cite{sivapuram2018topology}, etc, have been presented for topology optimization for different applications. Herein, we confine ourselves to density-based methods.
 
 TO problem in its original continuum form is well-known to be ill-posed, lacking closure of the solution space, and consequently, deficient in the existence of a solution \cite{Sigmund1998,allaire1993numerical}. This lack of existence of a solution is manifested through mesh-dependent results in the numerical framework. The numerical formulation of the problem, on the other hand, is closed and exhibits a solution, as the size of the elements governs the minimum feature size. A common approach to ensuring the existence of a solution is to restrict the solution space \cite{Eschenauer2001}. Over the years, various approaches have been proposed for the same. Ambrosio and Buttazo~\cite{ambrosio1993optimal} introduce the perimeter constraint method, which is numerically implemented by Haber et al.~\cite{Haber1996}. The perimeter constraint leads to mesh-independent solutions, but selecting the appropriate bound on the perimeter requires a certain level of experience. In addition, the method may result in infeasible solutions as it does not prevent the formation of thin members. Another readily implemented approach for imposing restrictions is not permitting rapid variations in the density field. Petersson and Sigmund~\cite{Petersson1998} introduce local constraints on the density gradients, capping the maximum density variation between adjacent elements. Sigmund~\cite{sigmund1997design} implements a sensitivity filter that modifies the sensitivities of the objective. Both local density gradient constraints and sensitivity filters lead to similar mesh-independent results, whereas sensitivity filtering is based on heuristics~\cite{sigmund1997design}.
 
 Bruns and Tortorelli~\cite{bruns2001topology} introduce density filtering, expressing the density of an element as the weighted sum of the element densities of neighboring elements. Weights in \cite{bruns2001topology} are calculated using a linear decaying function, while Bruns and Tortorelli~\cite{bruns2003element} and Wang and Wang~\cite{wang2005bilateral} evaluate weights using a Gaussian distribution function. Density filters are known to produce mesh-independent solutions. Bourdin~\cite{bourdin2001filters} proves the existence of a solution for a general case of density filtering. Solutions obtained using these methods exhibit gray regions, i.e., regions of intermediate densities, at the solid-void interface.
 
 Poulsen~\cite{Poulsen2003} implements a local length scale constraint that leads to many constraints. Various global length scale constraints have also been proposed within TO~\cite{Zhang2014,Guo2014,Xia2015,Singh2020}. Singh et al.~\cite{Singh2020} present an analytical problem relating to the permitted volume and minimum required length scale. The approach proposes a methodology to prevent convergence to undesired local minima when imposing length scale constraints. Guest et al.~\cite{Guest2004} propose the projection method that imposes length scales implicitly and yields close-to-black-and-white solutions. Mesh-independent solutions for the compliance minimization problem are presented. The projection method employs a parameter that provides the user implicit control over transition regions. This guarantees close-to-binary solutions, as one can always control the parameter and the length of transition regions. Guest~\cite{Guest2009a} extends the projection method to impose length scale on both solid and void phases. Sigmund~\cite{sigmund2007morphology} introduces morphological filters, producing close to 0-1 solutions. Wang et al.~\cite{wang2011projection} propose a robust formulation for getting close to 0-1 solutions while imposing a length scale on the solid phase using the projection filter, where a user selects the threshold and continuation strategy for the projection parameter. To the best of authors' knowledge, there is currently no density evaluation method that achieves the following properties without the use of parameters and weight functions: (i) ensures close-to-binary solutions with minimal transition regions, (ii) provides mesh-independent solutions, and (iii) is non-dependent on the user-defined density threshold, continuation approaches, or parameters.

 This paper introduces a novel density evaluation method using a scalar field product approach, termed the `\textit{normalized field product (nFP)}' method, which provides the properties mentioned above. The robustness and success of the nFP method are noted in solving various 2D and 3D  stiff structures and compliant mechanism design problems with their respective permitted resource constraint. For the former, compliance is minimized, whereas, for the latter, a multi-criteria objective is minimized~\cite{frecker_et_al_1994,saxena2000optimal}. The method is also extended with advanced elements~\cite{singh2024three} for 3D problems to subdue geometrical singularity and depict the proposed method's versatility. The new contributions of the manuscript are as follows:
 \let\labelitemi\labelitemii
 \begin{itemize}
 	\item A normalized field product-based approach for topology optimization to achieve close-to-binary optimal designs with embedded minimum length scale is proposed.
 	\item The introduced density evaluation is unique, parameter-free, imposes the desired length scale on solid phase, and is shown to yield close-to-binary solution without user intervention and without using heuristic-based continuation schemes.
 	\item The proposed density formulation is very general and allows the user to select appropriate functions for determining element density
 	\item The success and versatility of the approach are demonstrated by solving various benchmark 2D stiff structures and compliant mechanism problems. Numerical experiments are performed to assess mesh independence. Problems are solved with different volume fractions and length scales 
 	\item 
 	The method is readily extended to 3D cases. Stiff structure and compliant mechanism problems are solved using traditional hexahedral and advanced truncated octahedral elements.
 \end{itemize}
 
 The layout of the paper is structured as follows. Sec.~\ref{Sec:density_formulation} provides the novel density formulation. The definition of the product of a scalar field and its normalization are presented. Building upon the concepts, the novel density evaluation scheme is presented. The section also provides the gradient calculations and different density functions. Sec.~\ref{Sec:ProblemFormulation} notes the problem formulation wherein the optimization problems for stiff structures and compliant mechanisms are presented. The objectives sensitivity analyses are also reported. The numerical results and discussion are reported in Sec.~\ref{Sec:NumericalExamplesandDiscussion}. Mesh independent nature of the nFP approach is demonstrated for stiff structure and inverter mechanism problems. Solutions to problems with different volume fractions and length scales are reported. The method is extended to 3D, where stiff structure and compliant mechanism design problems are solved. The nFP method is also presented with truncated octahedron elements~\cite{singh2024three}. Lastly, conclusions are drawn in Sec.~\ref{Sec:Closure}.

\section{Density Formulation and Gradient calculation}\label{Sec:density_formulation}
This section first introduces the concepts and related terminologies of the product of a scalar field. Expanding on these concepts, a novel density modeling scheme is subsequently proposed. Various suitable functions are enlisted, and corresponding gradients are evaluated.	
\subsection{Product of a scalar field and its normalization}
Consider a scalar field $\gamma(\boldsymbol{\mathrm{X}}) > 0$, where $\boldsymbol{\mathrm{X}} \in \mathbb{R}$ is defined over a region $\mathbb{R}$. Let $P = {\mathbb{R}_1, \mathbb{R}_2,\ldots, \mathbb{R}_n}$ represent the $n$ partitions of $\mathbb{R}$. The product, denoted by $G$, of $\gamma(\boldsymbol{\mathrm{X}})$ over $\mathbb{R}$ is defined as \cite{riemann1867ueber}:
\begin{equation} \label{Eq:product_field}
	G = \lim\limits_{A_i \to 0}~ \prod_{i = 1}^{n} \gamma(\boldsymbol{\mathrm{X}}_i)^{A_i},
\end{equation}
where $\boldsymbol{\mathrm{X}}_i$ and $A_i$ are the centroid and area (2D)/volume (3D) of $\mathbb{R}_i$, respectively. The limit $A_i \to 0$ implies $n \to \infty$ and partition $\mathbb{R}_i$ converges to $\boldsymbol{\mathrm{X}}_i$.  With natural logarithm, Eq.~\ref{Eq:product_field} transpires to
\begin{eqnarray}\label{Eq:Prod_analytical_1}
	\ln G = \lim\limits_{ A_i \to 0}~ \sum\limits_{i=1}^{n} \ln (\gamma_i) A_i
	= \int\limits_{\mathbb{R}} \ln \gamma(\boldsymbol{\mathrm{X}}) d\mathbb{R}.
\end{eqnarray}
In view of Eq.~\ref{Eq:Prod_analytical_1}, one gets
\begin{eqnarray} \label{Eq:Prod_analytical}	
	G = \exp \left(\int\limits_{\mathbb{R}} \ln \gamma(\boldsymbol{\mathrm{X}}) d\mathbb{R} \right).
\end{eqnarray}
The analytical expression of $G$ (Eq.~\ref{Eq:Prod_analytical})	is termed in this paper as  \textit{Field Product} or \textit{FP} of  $\gamma(\boldsymbol{\mathrm{X}})$ over the region $\mathbb{R}$. A corollary of the above definition in the normalized sense can be written as
\begin{eqnarray}	\label{Eq:Field_prod}
	G_{nFP} = \lim\limits_{ A_i \to 0}~ \prod_{i = 1}^{n} \gamma(\boldsymbol{\mathrm{X}}_i)^{A_i/A(\mathbb{R})} = \exp \left( \dfrac{\displaystyle\int\limits_{\mathbb{R}} \ln \gamma(\boldsymbol{\mathrm{X}}) d\mathbb{R}}{\displaystyle\int\limits_{\mathbb{R}} d\mathbb{R}} \right),
\end{eqnarray} 
where {$G_{nFP}$ is called the \textit{normalized Field Product (nFP)} of $\gamma(\boldsymbol{\mathrm{X}})$ over $\mathbb{R}$,} and $A(\mathbb{R})$ denotes the area or volume of $\mathbb{R}$. Note that replacing $A_i$ with $A_i/A(\mathbb{R})$ non-dimensionalizes the exponent and normalizes the definition noted in Eq.~\ref{Eq:product_field}. In the next section, we use the expression obtained in Eq.~\ref{Eq:Field_prod} to define the new density modeling scheme. 

\subsection{Density evaluation scheme}
As mentioned above, TO methods~\cite{bendsoe2003topology} determine the optimal density distribution $0 \leq \rho(\boldsymbol{\mathrm{X}}) \leq 1$ within the given design domain $\Omega$ while extremizing the desired objective function with prescribed constraints (if any). Let the auxiliary field, $\alpha(\boldsymbol{\mathrm{X}})$, where $\boldsymbol{\mathrm{X}} \in \Omega$, be a scalar field defined within the design domain. The motive is to formulate the density field, $\rho(\boldsymbol{\mathrm{X}})$, to $\alpha(\boldsymbol{\mathrm{X}})$ and solve for $\alpha(\boldsymbol{\mathrm{X}})$ as the primary variable. The way one relates $\rho(\boldsymbol{\mathrm{X}})$ in terms of  $\alpha(\boldsymbol{\mathrm{X}})$ furnishes different density evaluations.  For instance, density filtering~\cite{bruns2001topology} represents $\rho(\boldsymbol{\mathrm{X}})$ as a convolution integral between the auxiliary field $\alpha(\boldsymbol{\mathrm{X}}^*)$ and a user-defined weight function, $w(\boldsymbol{\mathrm{X}},\boldsymbol{\mathrm{X}}^*)$ over the neighborhood of $\boldsymbol{\mathrm{X}}$ (a finite region around $\boldsymbol{\mathrm{X}}$), $\Gamma(\boldsymbol{\mathrm{X}})$, that is, $\boldsymbol{\mathrm{X}}^* \in \Gamma(\boldsymbol{\mathrm{X}})$. In contrast, projection \cite{Guest2004} exponentiates a scaled version of the convolution integral. The selection of expression allows for the implicit imposition of restrictions on the nature of $\rho(\boldsymbol{\mathrm{X}})$. Note that $\Gamma(\boldsymbol{\mathrm{X}})$  indicates the neighborhood of $\boldsymbol{\mathrm{X}}$, i.e., a finite region surrounding $\boldsymbol{\mathrm{X}}$ depending upon the the user-defined shapes~\cite{svanberg2013density}. In this context, we establish a novel relationship between the two to facilitate close to transition-free topologies with a minimum length scale on the solid phase without the need for additional parameters.

In a typical TO setting, the associated PDEs are solved using the finite element methods. The design domain $\Omega$ is parameterized using $n$ finite elements $\Omega_i|_{i = 1, \,2,\, 3,\,\cdots,\,n}$ having centroids $\mathbf{X}_i$.  Herein, we designate element $i$ with a density variable $\rho_i$ and a scalar variable  $\alpha_i$. These variables are considered constant within the element. The descretized domain has  $\boldsymbol{\rho} = \{\rho_1,~\rho_2, \hdots, \rho_n\}$ and $\boldsymbol{\alpha} = \{\alpha_1,~\alpha_2, \hdots, \alpha_n\}$ as the density and scalar vectors, respectively. Incorporating the length scale within the formulation requires information on the neighborhood of each element. Let $\mathbb{N_i}$ be the set of elements that lie within the neighborhood of element $i$, $\Gamma_i \equiv \Gamma(\boldsymbol{\mathrm{X}}_i)$, that is, $\mathbb{N}_i = \{j \mid \boldsymbol{\mathrm{X}}_j \in \Gamma_i\}$~\cite{svanberg2013density}. 

To achieve close-to-binary solutions with minimum length scale on solid phase, we develop an expression such that the density in element $i$ is 1 if any element in $\Gamma_i$ has auxiliary field value as 1, that is, $\rho_i = 1$ if $\alpha_j = 1$ for any $j \in \mathbb{N}_i$ Mathematically, one can write it as

\begin{eqnarray}	\label{Eq:BPP}
	\rho_i = 1 - \prod_{j \in \mathbb{N}_i} (1 - \alpha_j),
\end{eqnarray}
where $0 \leq \alpha_j \leq 1$. The relation in Eq.~\ref{Eq:BPP} can provide optimized solutions close to 0-1; however, the solutions will be mesh-dependent. Say, the auxiliary variable for each element is $\alpha_0$ and $\mathbb{N}_i$ contains $r$ finite elements. Per Eq.~\ref{Eq:BPP}, one writes $\rho_i = 1 - (1- \alpha_0)^r$, which varies with mesh alteration; thus, optimized solutions will be mesh dependent, which is undesirable. The normalized field product concepts discussed above are employed to evaluate $\boldsymbol{\rho}$ independent of the mesh. In view of Eq.~\ref{Eq:Field_prod} and Eq.~\ref{Eq:BPP},  one writes the density of element $i$, $\rho_i$ as

\begin{eqnarray}	\label{Eq:BPP_mod}
	\rho_i = 1 - \prod_{j \in \mathbb{N}_i} (1 - \alpha_j)^\frac{A(\Omega_j)}{A(\Gamma_i)},
\end{eqnarray}
where $A(\Gamma_i)=\sum\limits_{j \in \mathbb{N}_i} A(\Omega_j)$.
\begin{figure}[!ht]
	\begin{subfigure}[t]{0.45\textwidth}
		\centering
		\includegraphics[scale = 0.75]{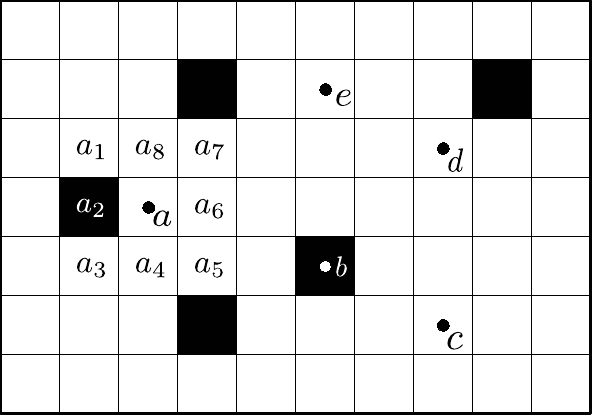}
		\caption{}
		\label{fig:density1}
	\end{subfigure}
	\begin{subfigure}[t]{0.45\textwidth}
		\centering
		\includegraphics[scale = 0.75]{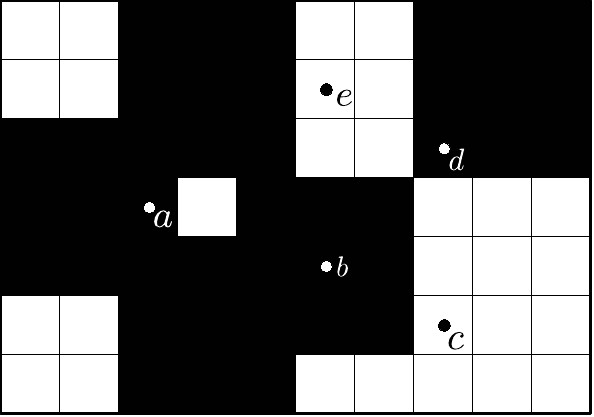}
		\caption{}
		\label{fig:density2}
	\end{subfigure}
	\caption{A schematic diagram for the proposed density formulation. (\subref{fig:density1}) elements with $\alpha =1$ are shown in black, whereas remnant elements are with $\alpha =0$.  (\subref{fig:density2})  Suppose a set of neighborhood elements contains only the immediate elements. For example, elements $a_1,\,a_2,\,\cdots,\,a_8$ are the neighbor elements for element $a$. As element $a_2$ has, $\alpha =1$, $\rho_i|_{i= a,\,a_1,\,a_2,\,a_3,\,a_4,\,a_5,\,a_7,\, a_8} =1$, whereas $\rho_i|_{i= a_6} =0$ as per Eq.~\ref{Eq:BPP_mod}. Likewise, one determines the density variable of other elements and plots the final density as shown in the right-side figure.}	\label{fig:Densitydemonstration}
\end{figure}
\noindent A simple working schematic diagram of the formulation is demonstrated in Fig.~\ref{fig:Densitydemonstration}. Consider a design domain discretized using rectangular elements, wherein elements with $\alpha =1$ are depicted in black, and the remaining elements are with $\alpha =0$. Say, $\mathbb{N}$, a set of neighborhood elements for each element contains only the immediate neighbor of the element. For example, elements $a_1,\,a_2,\,\cdots,\,a_8$ are the neighbor elements for element $a$ (Fig.~\ref{fig:density1}). Using (Eq.~\ref{Eq:BPP_mod}),  $\rho_i$ is plotted in Fig.~\ref{fig:density2}. The analytic expression of $\boldsymbol{\rho}$ is obtained by replacing $\gamma(\boldsymbol{\mathrm{X}})$ by  $(1 - \alpha(\boldsymbol{\mathrm{X}}))$ in Eq.~\ref{Eq:Field_prod}  as
\begin{equation}
	\rho(\boldsymbol{\mathrm{X}}) = 1 - \exp \left( \dfrac{\displaystyle\int\limits_{\Gamma(\boldsymbol{\mathrm{X}})} \ln (1- \alpha(\boldsymbol{\mathrm{X}})) dA}{\displaystyle\int\limits_{\Gamma(\boldsymbol{\mathrm{X}})} dA} \right).
\end{equation}
With $1- \alpha(\boldsymbol{\mathrm{X}}) = f(\beta(\boldsymbol{\mathrm{X}}))$, where $f(.)$ is an invertible function and $0 < f(x) \leq 1$ with further restrictions mentioned later, Eq.~\ref{Eq:BPP_mod} transpires to
\begin{eqnarray}\label{Eq:nFP_density_beta}
	\rho_i = 1 - \prod_{j \in \mathbb{N}_i} f(\beta_j)^\frac{A(\Omega_j)}{{A(\Gamma_i)}},
\end{eqnarray}
$\boldsymbol{\beta} = \{\beta_1,~\beta_2,~ \hdots, \beta_{n}\}$ is the element-wise constant approximation of $\beta(\boldsymbol{\mathrm{X}})$ within the given design domain, i.e., $\beta_i$ is a \textit{local} variable. Herein the objective is to determine $\beta_i$; thus $\rho_i$ for element~$i$ during the optimization process. Irrespective of the function choice, $f(.)$, which satisfies the above-mentioned properties, the density evaluation in Eq.~\ref{Eq:nFP_density_beta}  furnishes the following desirable properties:
\begin{itemize}
	\item  Permits close to 0-1 optimized designs,
	\item  Does not need user-defined parameters
	\item Provides discretization independent evaluation of densities by virtue of the expression noted in Eq.~\ref{Eq:nFP_density_beta} and
	\item Implicitly imposes length scale on the density distribution on the solid phase via the defined neighborhood $\mathbb{N}_i$.
\end{itemize}
\begin{table}[h!]
	\centering
	\begin{tabular}{c|c|c|c|c}
		\hline
		\textbf{Choice no} & \textbf{$f(\beta_j)$}                  & \textbf{Range of} $\beta_j$ & \textbf{$T_2=\dfrac{1}{f(\beta_j)}  \dfrac{d f(\beta_j)}{d \beta_j} $} & \textbf{Bound for} $T_2$ \\ \hline
		1                & $e^{\beta_j}$                          & $(-\infty,0]$                        & 1                                                               & 1                 \\ \hline
		2                & $1 - \tanh \beta_j$                    & $[0,\infty)$                        & $-(1 + \tanh \beta_j)$                                                                  & $(-2,\, -1]$                  \\ \hline
		3                & $\dfrac{1}{\beta_j^n}$                 & $[1,\infty)$                        & $-\dfrac{n}{\beta_j}$                                                                  & $[-n,\,0)$                 \\ \hline
		4                & $1 - \dfrac{\tan^{-1} \beta_j}{\pi/2}$ & $[0,\infty)$                         & $-\dfrac{1}{ \left(\pi/2 - \tan^{-1}\beta_j \right) \left( 1 + \beta_j^2\right)}$                                                                 & $[-1,\,0]$                 \\ \hline
	\end{tabular}
	\caption{List of some possible $f(\beta_j)$ and their contribution to density gradients.}	\label{Tab:fucntion_choices}
\end{table}

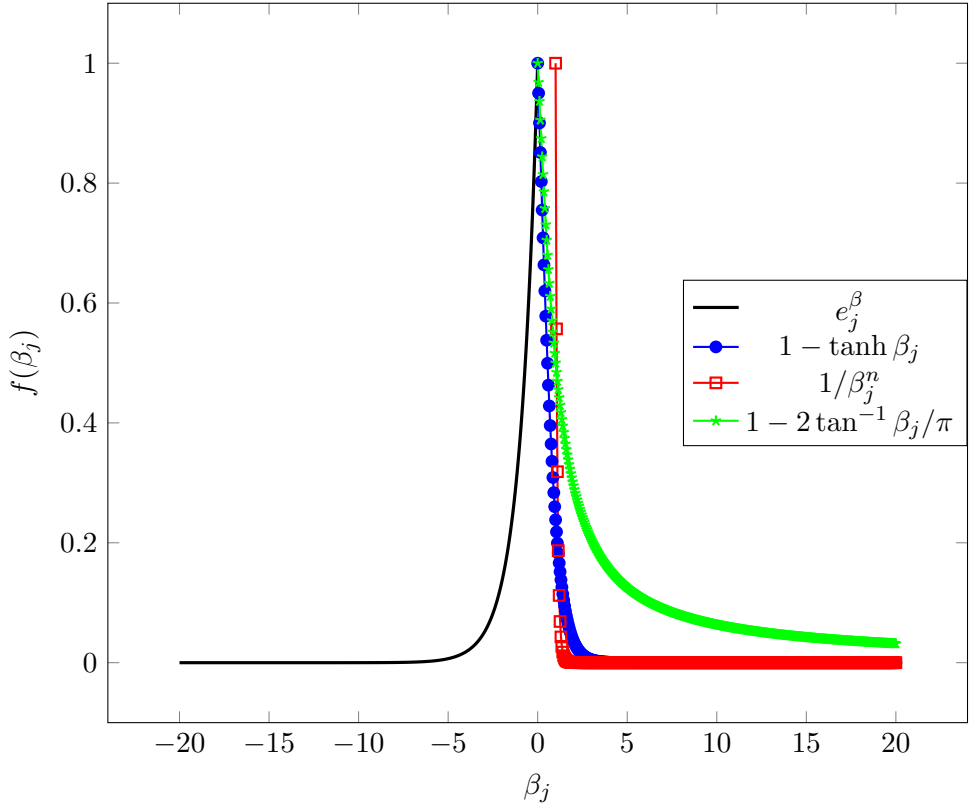
\begin{figure*}[h] 
	\centering
	\begin{tikzpicture} 
		\pgfplotsset{compat =1.9}
		\begin{axis}[
			width = 0.75\textwidth,
			xlabel= $\beta_j$,
			ylabel= $f(\beta_j)$,
			legend style={at={(1,0.5)},anchor=east}]
			\pgfplotstableread{data1.txt}\mydata;
			\addplot[black, mark size=1pt,style={very thick}]
			table {\mydata};
			\addlegendentry{$e^\beta_j$}
			\pgfplotstableread{data2.txt}\mydata;
			\addplot[blue,mark = *,mark size=2pt,style= thick]
			table {\mydata};
			\addlegendentry{$1 - \tanh \beta_j$}
			\pgfplotstableread{data3.txt}\mydata;
			\addplot[red,mark = square,mark size=2pt,style= thick]
			table {\mydata};
			\addlegendentry{${1/\beta_j^n}$}
			\pgfplotstableread{data4.txt}\mydata;
			\addplot[green,mark = star,mark size=2pt,style= thick]
			table {\mydata};
			\addlegendentry{$1 - 2\tan^{-1} \beta_j/\pi$}
		\end{axis}
	\end{tikzpicture}
	\caption{$f(\beta_j)$ plots. $n=12$ is picked for the third function.} \label{fig:function_plot}
\end{figure*}
\subsection{Gradient calculation and $f(\beta)$ selection}
We use gradient-based optimization in the approach. Therefore, we need to determine the gradients of the objective(s) and constraint(s) with respect to the design variable $\beta_j$. Consider generic function $f_{0}$ that indicates objective or constraint. Gradient of  $f_{0}$ is determined using the chain rule since density variable  $\rho_i$ depends upon design variable $\beta_j$ (Eq.~\ref{Eq:nFP_density_beta}) as: 
\begin{equation}\label{Eq:Chain_rule}
	\frac{\partial f_0}{\partial \beta_j}  = \sum_{i\in\mathbb{N}_j} \frac{\partial f_0}{\partial \rho_i}\frac{\partial \rho_i}{\partial \beta_j}.
\end{equation}
The procedure to determine $\frac{\partial f_0}{\partial \rho_i}$ is outlined in Sec.~\ref{Sec:Sensitivityanalysis}. Herein, we  determine $\frac{\partial \rho_i}{\partial \beta_j}$ as
\begin{eqnarray} 	\label{Eq:density_grad}
	\dfrac{\partial\rho_i}{\partial \beta_j} = 
	\begin{cases}
		-\dfrac{(1-\rho_i)A(\Omega_j)}{{A(\Gamma_i)}}  \left( \dfrac{1}{f(\beta_j)} \dfrac{d f(\beta_j)}{d \beta_j} \right)  &\text{for}~ i \in \mathbb{N}_j\\
		0 &\text{for}~ i \notin \mathbb{N}_j.
	\end{cases}
\end{eqnarray}
To ensure the gradients are not singular (Eq.~\ref{Eq:density_grad}), $\dfrac{1}{f(\beta_j)}\dfrac{d f(\beta_j)}{d \beta_j} $ must be finite. Both conditions can be met by functions that asymptotically approach 0. Table \ref{Tab:fucntion_choices} provides some such functions. Fig.~\ref{fig:function_plot} indicates their plot, wherein one can note that these functions lie between [0,\,1], have monotonous behavior, and asymptotically tend to 0. The last property indicates that the density variables asymptotically advance towards 1 (Eq.~\ref{Eq:nFP_density_beta}).   

\section{Problem formulation}\label{Sec:ProblemFormulation}
We use the SIMP material model \cite{bendsoe1999material}
for which the elemental stiffness, $\boldsymbol{\mathrm{Ke}}$, for an element with density $\rho_i$ is given as,
\begin{eqnarray}\label{Eq:SIMP}
	\boldsymbol{\mathrm{Ke}} = \{\rho_i^\eta (1-\rho_{min}) + \rho_{min}\}\boldsymbol{\mathrm{K}}_0,
\end{eqnarray}
where $\eta$ is the SIMP penalty parameter, $\rho_{min}$ is a small positive number introduced to remove potential singularity of the stiffness matrix, and $\boldsymbol{\mathrm{K}}_0$ is the elemental stiffness of a solid element~\cite{bendsoe2003topology}. In the following subsections, we provide the optimization problem formulation for the structure and compliant mechanisms (CMs),  the sensitivities of the objectives with respect to the design vector $\bm{\beta}$ using the chain rule (Eq.~\ref{Eq:Chain_rule}) and the adjoint-variable method.

\subsection{Stiff structures}
The presented approach is demonstrated by solving 2D and 3D stiff structure design problems. The conventional formulation, namely minimizing compliance or strain energy, is employed for designing the stiff structures. In this context, the optimization problem is expressed as follows:
\begin{equation}\label{Eq:OptimizationequationStiff}
	\begin{rcases}
		& \underset{\bm{\rho(\beta)}}{\text{min}}
		& &{f_0(\bm{\rho(\beta)})} = \mu_s\frac{1}{2} \mathbf{u}^\top \mathbf{Ku} = \mu_s SE\\
		& \text{such that:}  &&\,\, \mathbf{Ku} = \mathbf{F}\\
		&  && \,\, \text{g}_1=\frac{\sum_{i=1}^{n}\rho_i(\bm{\beta})V{_i}}{V^*}-1\le 0\\
		&  && \beta_{l} \le \beta_i \le \beta_u
	\end{rcases},
\end{equation}
where $f_0(\bm{\rho(\beta)})$ is the objective function to be optimized. $SE$ indicates the strain energy. $\mathbf{K}$ and $\mathbf{u}$ represent the global stiffness matrix and displacement vector, respectively. $\mathbf{F}$ is the external force vector. $\mu_s$, a scaling factor, is employed primarily to adjust the magnitude of the objective; thus, sensitivities consistently. $V^*$ is the permitted volume of the designs. $V_i$ is the volume of the element~$i$. $\beta_l$ and $\beta_u$ are the lower and upper bounds on the design variables. 


\subsection{Compliant mechanisms}
To demonstrate  diversity of the proposed nFP approach, we also design 2D and 3D Compliant mechanisms (CMs) using this approach. Such mechanisms have monolithic designs that transfer/transform force, motion, or energy into the desired work. For designing such mechanisms using TO, typically, an objective stemming from a flexibility measure (e.g., output deformation) and a stiffness measure (e.g., strain energy) is optimized~\cite{frecker_et_al_1994,saxena2000optimal}. The former provides the desired output deformation, whereas the latter helps mechanisms to sustain under the applied loads. We use a multi-criteria  objective~\cite{frecker_et_al_1994,saxena2000optimal}, based on the above two  measures of CMs. We solve the following optimization problem for designing the CMs:

\begin{equation}\label{Eq:OptimizationequationCMs}
	\begin{rcases}
		& \underset{\bm{\rho(\beta)}}{\text{min}}
		& &{f_0(\bm{\rho(\beta)})} = -\mu_{CM}\frac{\mathbf{v}^\top \mathbf{K}\mathbf{u}}{\mathbf{u}^\top \mathbf{K}\mathbf{u}} = -\mu_{CM} \frac{MSE}{2SE}\\
		& \text{such that:}  &&\,\, \mathbf{Ku} = \mathbf{F}\\
		&  &&\,\, \mathbf{Kv = F_\mathrm{d}}\\
		&  && \,\, \text{g}_1=\frac{\sum_{i=1}^{n}\rho_i(\bm{\beta})V{_i}}{V^*}-1\le 0\\
		&  && \beta_{l} \le \beta_i \le \beta_u
	\end{rcases},
\end{equation}
where $\mb{F}_\text{d}$ is the unit dummy force applied in the direction of the desired deformation at the output location of the CMs~\cite{saxena2000optimal}. $MSE$ is called the mutual strain energy~\cite{saxena2000optimal}. $\mathbf{v} $ is obtained displacement vector corresponding to $\mb{F}_\text{d}$. $\mu_{CM}$ is an analogous to $\mu_s$ for CMs. In Eq.~\ref{Eq:OptimizationequationCMs}, the remaining parameters have the same meaning as mentioned in Eq.~\ref{Eq:OptimizationequationStiff}. All mechanical balanced equations noted in Eqs.~\ref{Eq:OptimizationequationStiff} and \ref{Eq:OptimizationequationCMs} are solved with small deformation finite element analysis assumptions.

\subsection{Sensitivity analysis} \label{Sec:Sensitivityanalysis}
We use the gradient-based optimizer, the MMA~\cite{svanberg1987method}, for topology optimization. The adjoint-variable method is employed to determine the sensitivities of the objectives and constraints with respect to the density variable $\bm{\rho}$. An aggregate performance function $\mathcal{L}$ for evaluating the sensitivities can be written as:

\begin{equation}\label{Eq:performancefunction}
	\begin{aligned}
		{\mathcal{L} (\mathbf{u},\mathbf{v},\,\bm{\rho})} = f_0(\mathbf{u},\mathbf{v},\bm{\rho}) + \trr{\bm{\lambda}}_1 \left(\mathbf{Ku -F}\right) + \trr{\bm{\lambda}}_2 (\mathbf{Kv-F_\mathrm{d}}),
	\end{aligned}
\end{equation}
where  $\bm{\lambda}_1$ and $\bm{\lambda_2}$ are the Lagrange multipliers. The sensitivity with respect to the density variable can be written as
\begin{equation}
	\frac{d f_0}{d \rho_i} = \frac{\partial f_0}{\partial \rho_i} + \underbrace{\left(\frac{\partial f_0}{\partial \mathbf{u}} + \trr{\bm{\lambda}}_1 \mathbf{K}\right)}_{\text{Term1}} \frac{\partial \mathbf{u}}{\partial \rho_i} + \trr{\bm{\lambda}}_1 \frac{\partial \mathbf{K}}{\partial \rho_i}\mathbf{u} + \underbrace{\left(\frac{\partial f_0}{\partial \mathbf{v}} + \trr{\bm{\lambda}}_2 \mathbf{K}\right)}_{\text{Term2}} \frac{\partial \mathbf{v}}{\partial \rho_i} + \trr{\bm{\lambda}}_2 \frac{\partial \mathbf{K}}{\partial \rho_i}\mathbf{v},
\end{equation}
$\bm{\lambda}_1$ and $\bm{\lambda_2}$ are selected such that Term1 and Term2 vanish, i.e.,

\begin{equation}\label{Eq:lagrangemultipliers}
	\begin{rcases}
		\trr{\bm{\lambda}}_1  &= -\pd{f_0(\mathbf{u},\, \mathbf{v},\,\bm{\rho})}{\mathbf{u}} \inv{\mathbf{K}}\\
		\trr{\bm{\lambda}}_2  &= -\pd{f_0(\mathbf{u},\, \mathbf{v},\,\bm{\rho})}{\mathbf{v}} \inv{\mathbf{K}}
	\end{rcases}.
\end{equation}
Using the above multipliers (Eq.~\ref{Eq:lagrangemultipliers}), one   evaluates sensitivities of the objective with respect to $\rho_i$ as

\begin{equation}\label{Eq:objectivesensitivity}
	\frac{d f_0}{d \rho_i} = \pd{f_0}{\rho_i} + \trr{\bm{\lambda}}_1\pd{\mathbf{K}}{\rho_i}\mathbf{u} +  \trr{\bm{\lambda}}_2\pd{\mathbf{K}}{\rho_i}\mathbf{v},
\end{equation}

\subsubsection{Stiff structures}
In the case of stiff structure designs, we discard the variables $\mathbf{v}$ and Lagrange multiplier $\lambda_2$. With $f_0 = \mu_s\frac{1}{2} \mathbf{u}^\top \mathbf{Ku}$ as strain energy (Eq.~\ref{Eq:OptimizationequationStiff}), Eq.~\ref{Eq:lagrangemultipliers} yields to $\lambda_1^\top =-2\mathbf{u}^\top$.  Now in view of $\lambda_1$, Eq.~\ref{Eq:objectivesensitivity} transpires to
\begin{equation}\label{Eq:ComplianceSens}
	\frac{\partial f_0}{\partial \rho_i} =  -\mu_s\frac{1}{2}\trr{\mathbf{u}}\pd{\mathbf{K}}{\rho_i}\mathbf{u} = 	- \mu_s\frac{\eta (1-\rho_{min})}{2} \rho^{\eta -1}_i \boldsymbol{\mathrm{u}}^\top \boldsymbol{\mathrm{K}}_0\boldsymbol{\mathrm{u}}.
\end{equation}
Using Eqs.~\ref{Eq:density_grad} and \ref{Eq:ComplianceSens} in Eq.~\ref{Eq:Chain_rule}, one can determine $\frac{\partial f_0}{\partial \beta_j}$ for the stiff structure designs.

\subsubsection{Compliant mechanisms}
To design CMs we need adjoint variables $\lambda_1$ and $\lambda_2$ for determining the objective sensitivities. With multi-criteria objective~\cite{saxena2000optimal}, i.e., $f_0 = -\mu_{CM}\frac{\mathbf{v}^\top \mathbf{K}\mathbf{u}}{\mathbf{u}^\top \mathbf{K}\mathbf{u}}$, Eq.~\ref{Eq:lagrangemultipliers} yields

\begin{equation}\label{Eq:CMs_lambdas}
	\begin{rcases}
		\trr{\bm{\lambda}}_1 & = \mu_{CM}\left(\frac{1}{\mathbf{u}^\top \mathbf{Ku}} \mathbf{v}^\top -\frac{\mathbf{v}^\top \mathbf{Ku}}{{(\mathbf{u}^\top \mathbf{K}\mathbf{u} )}^2}2\mathbf{u}^\top\right)\\
		\trr{\bm{\lambda}}_2 &= \mu_{CM}\left(\frac{1}{\mathbf{u}^\top \mathbf{Ku}} \mathbf{u}^\top\right)
	\end{rcases}.
\end{equation}
In view of Eq.~\ref{Eq:CMs_lambdas}, the objective sensitivities for CM designs with respect to $\rho_i$ now can be determined as
\begin{equation}\label{Eq:CMSens}
	\frac{\partial f_0}{\partial \rho_i} =   \mu_{CM}\left[\frac{\mathbf{v}^\top \mathbf{Ku}}{{(\mathbf{u}^\top \mathbf{K}\mathbf{u} )}^2}\left(-\trr{\mathbf{u}}\pd{\mathbf{K}}{\rho_i}\mathbf{u}\right) + \frac{1}{\mathbf{u}^\top \mathbf{Ku}}\left(\trr{\mathbf{u}}\pd{\mathbf{K}}{\rho_i}\mathbf{v}\right)\right].
\end{equation}
One can use Eqs.~\ref{Eq:density_grad} and \ref{Eq:CMSens} in Eq.~\ref{Eq:Chain_rule} to determine $\frac{\partial f_0}{\partial \beta_j}$ for optimizing the CMs. We add a spring with stiffness $K_s$ representing the workpiece stiffness at the output location~\cite{saxena2000optimal}. The spring motivates the optimizer to connect the input and output ports with sufficient material.

\begin{figure}[h!]
	\begin{subfigure}[t]{0.20\textwidth}
		\centering
		\includegraphics[scale = 0.5]{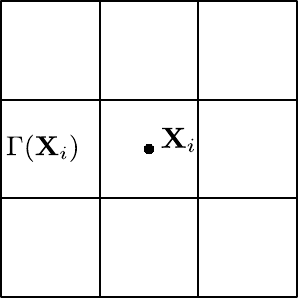}
		\caption{$ls =1$}
		\label{fig:ls=1}
	\end{subfigure}
	\begin{subfigure}[t]{0.20\textwidth}
		\centering
		\includegraphics[scale = 0.5]{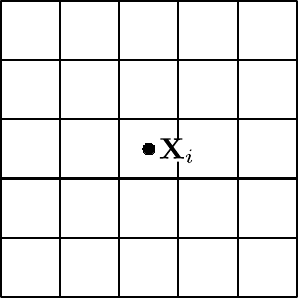}
		\caption{$ls=2$}
		\label{fig:ls=2}
	\end{subfigure}
	\begin{subfigure}[t]{0.20\textwidth}
		\centering
		\includegraphics[scale = 0.5]{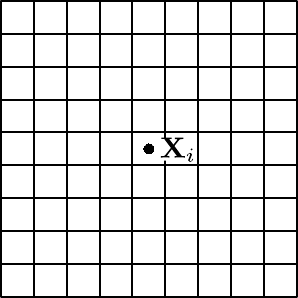}
		\caption{$ls=4$}
		\label{fig:ls=4}
	\end{subfigure}
	\begin{subfigure}[t]{0.35\textwidth}
		\centering
		\includegraphics[scale = 0.55]{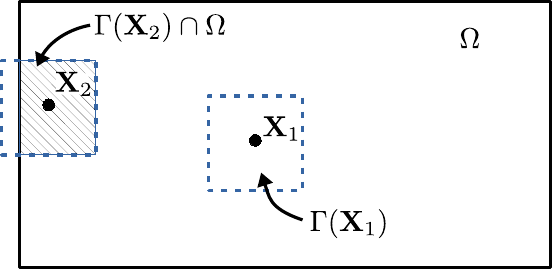}
		\caption{}
		\label{fig:Boundary_neighborhood}
	\end{subfigure}
	\caption{\subref{fig:ls=1}, \subref{fig:ls=2} and \subref{fig:ls=4} depict discretization of the neighborhood $\Gamma(\boldsymbol{\mathrm{X}})$ for an interior point at different levels of refinement. (\subref{fig:Boundary_neighborhood}) indicates interaction of neighborhood with domain for points close to the domain boundary.}\label{fig:Lenght_Neighbourhood_points}
\end{figure}
\section{Numerical results and Discussion} \label{Sec:NumericalExamplesandDiscussion}
This section provides design optimization of various 2D and 3D stiff structures and compliant mechanisms to demonstrate the effectiveness and versatility of the proposed nFP approach. First, we show optimization for 2D stiff structures and CMs. The success of the approach for these problems is presented given (a) mesh independence, (b) natural tendency for providing close-to binary solutions, i.e., low grayness measure defined as~\cite{sigmund2007morphology}
\begin{equation}\label{Eq:grayness_measure}
	g(\bm{\rho}) = \sum\limits_{i = 1}^{n} \frac{4\rho_i(1-\rho_i)}{n}.
\end{equation}
In addition, for compliance minimization and multi-criteria objective optimization, the method's capacity to generate a range of topologies is examined for various problem specifications like volume fractions and length scale.

\begin{figure}
	\begin{subfigure}{0.45\textwidth}
		\centering
		\includegraphics[scale = 0.5]{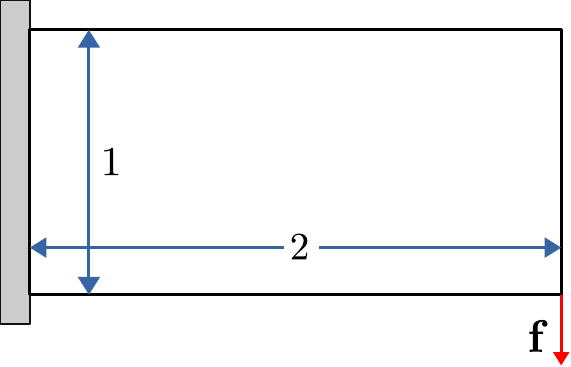}
		\caption{Cantilever beam design}
		\label{fig:Prob_description_Canti}
	\end{subfigure}
	\,
	\begin{subfigure}{0.45\textwidth}
		\centering
		\includegraphics[scale = 0.5]{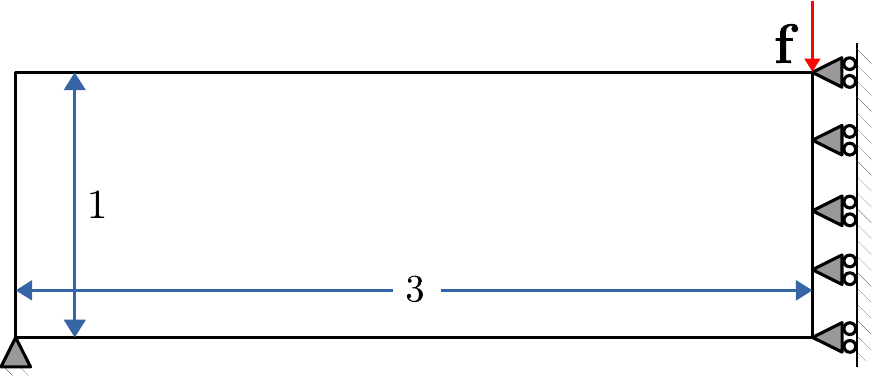}
		\caption{Mid-Load beam design}
		\label{fig:Prob_description_MLB}
	\end{subfigure}
	\caption{Problem description for stiff structures}\label{fig:Prob_description_stiff}
\end{figure}
\begin{figure}[!htb]
	\centering
	\includegraphics[width=0.35\textwidth]{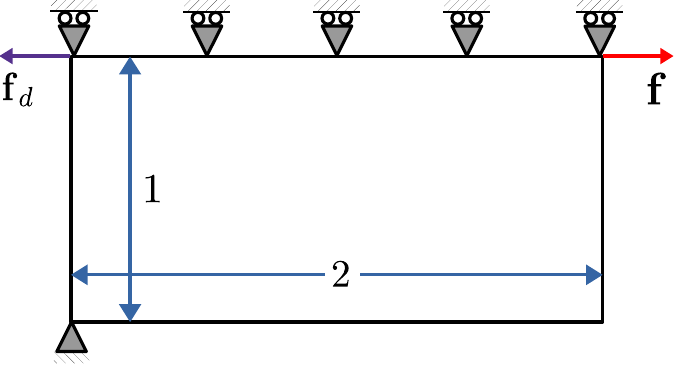}
	\caption{Problem description for a displacement inverter.}
	\label{fig:Prob_description_disp_invert}
\end{figure}

\begin{figure}[h!]
	\centering
	\begin{subfigure}{0.30\textwidth}
		\centering
		\includegraphics[scale =0.37]{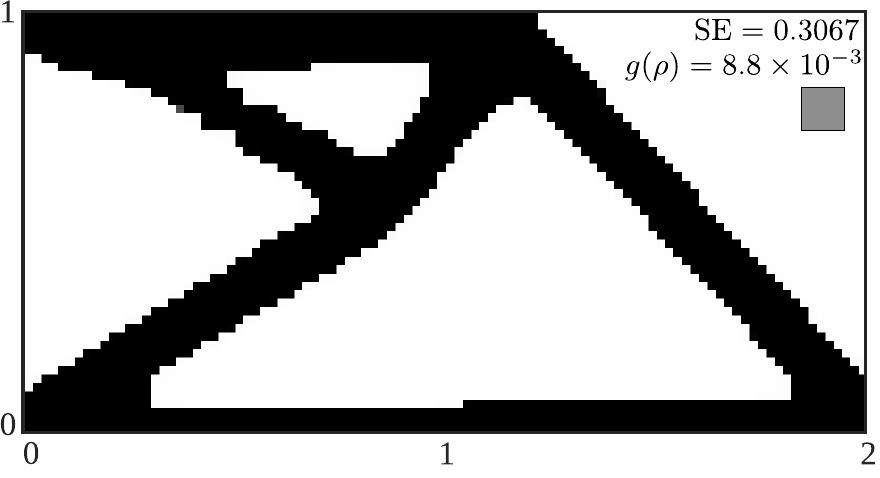}
		\caption{Mesh: $100\times 50$, $v_f = 0.35$, $ls = 2$} \label{fig:Sol_Canti_100x50_vf_35_ls_2}%
	\end{subfigure}
	\,
	\begin{subfigure}{0.30\textwidth}
		\centering
		\includegraphics[scale =0.37]{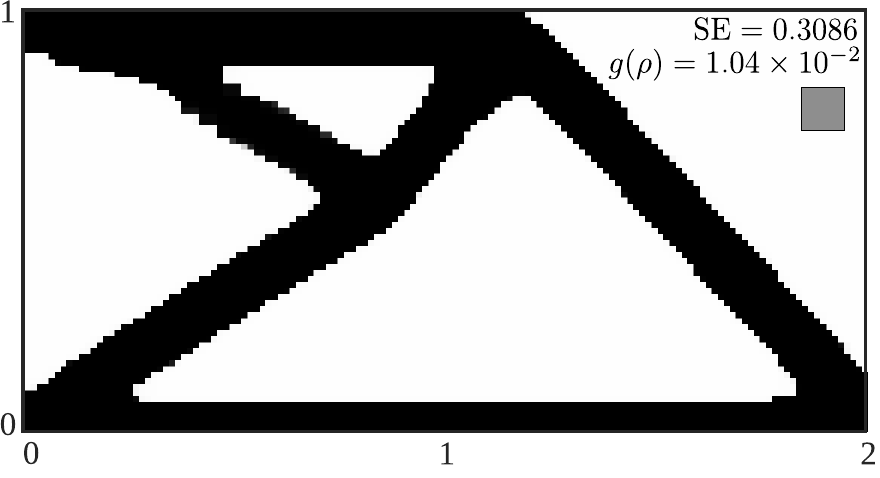}
		\caption{Mesh: $140\times 70$, $v_f = 0.35$, $ls = 3$} 
		\label{fig:Sol_Canti_140x70_vf_35_ls_3}%
	\end{subfigure}
	\,
	\begin{subfigure}{0.30\textwidth}
		\centering
		\includegraphics[scale =0.37]{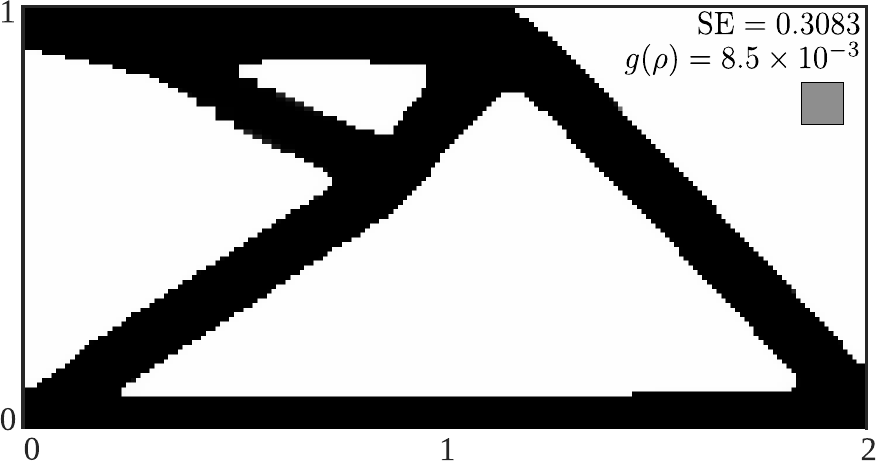}
		\caption{Mesh: $180\times 90$, $v_f = 0.35$, $ls = 4$}
		\label{fig:Sol_Canti_180x90_vf_35_ls_4}%
	\end{subfigure}
	\,
	\begin{subfigure}{0.30\textwidth}
		\centering
		\includegraphics[scale =0.36]{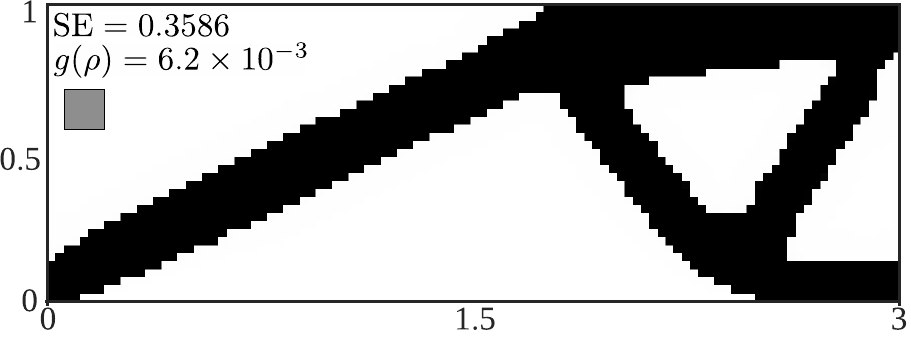}
		\caption{Mesh: $105\times 35$, $v_f = 0.35$, $ls = 2$ }
		\label{fig:Sol_MBB_105x35_vf_35_ls_2}%
	\end{subfigure}
	\,
	\begin{subfigure}{0.33\textwidth}
		\centering
		\includegraphics[scale =0.35]{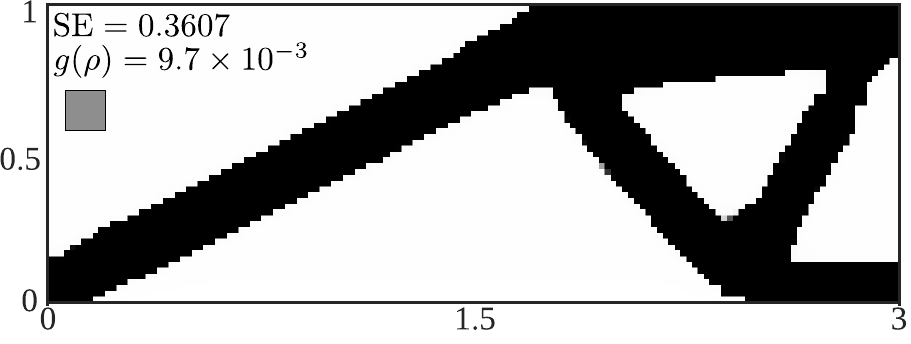}
		\caption{Mesh: $147\times 49, v_f = 0.35, ls = 3$ }
		\label{fig:Sol_MBB_147x49_vf_35_ls_3}%
	\end{subfigure}
	\,
	\begin{subfigure}{0.33\textwidth}
		\centering
		\includegraphics[scale =0.35]{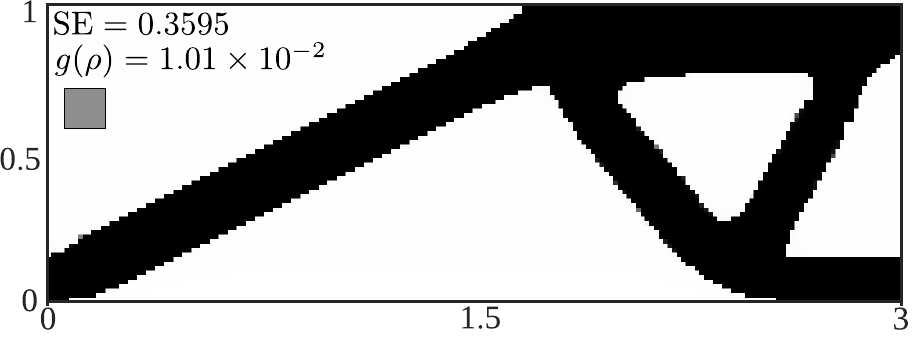}
		\caption{Mesh: $189\times 63, v_f = 0.35, ls = 4$ }
		\label{fig:Sol_MBB_189x63_vf_35_ls_4}%
	\end{subfigure}
	\,
	\begin{subfigure}{0.3\textwidth}
		\centering
		\includegraphics[scale =0.37]{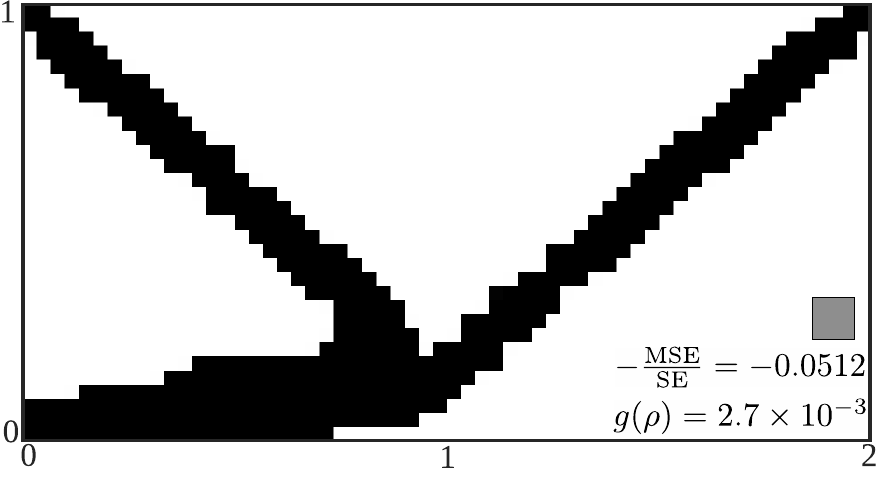}
		\caption{Mesh: $60\times 30$, $v_f = 0.22$, $ls = 1$}
		\label{fig:Sol_Disp_inverter_60x30_vf_22_ls_1}%
	\end{subfigure}
	\,
	\begin{subfigure}{0.3\textwidth}
		\centering
		\includegraphics[scale = 0.37]{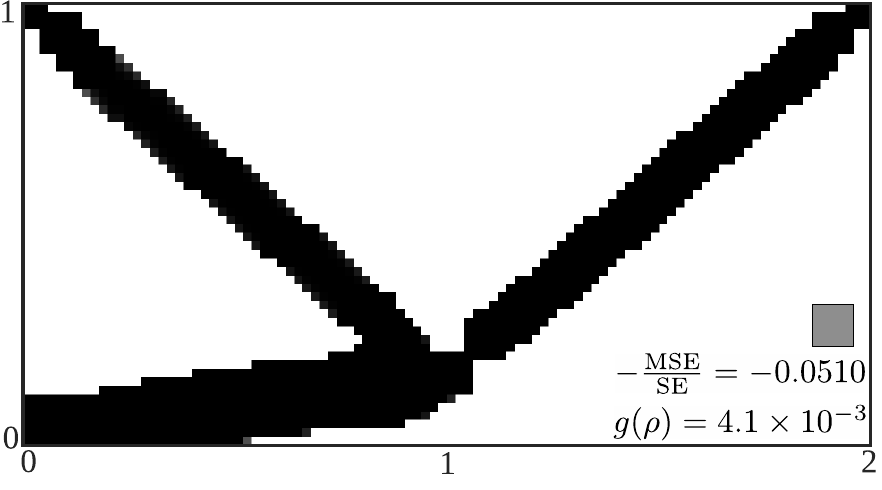}
		\caption{Mesh: $100\times 50$, $v_f = 0.22$, $ls = 2$}
		\label{fig:Sol_Disp_inverter_100x50_vf_22_ls_2}%
	\end{subfigure}
	\begin{subfigure}{0.3\textwidth}
		\centering
		\includegraphics[scale = 0.37]{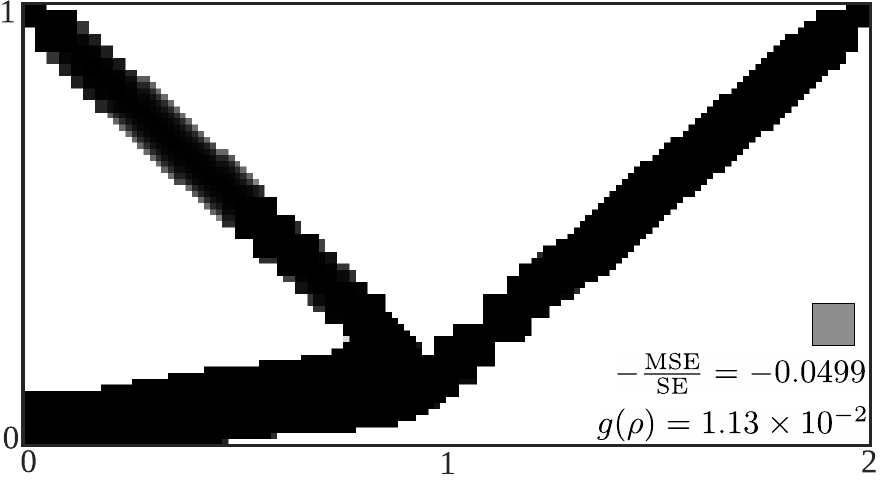}
		\caption{Mesh: $140\times 70$, $v_f = 0.22$, $ls = 3$}
		\label{fig:Sol_Disp_inverter_140x70_vf_22_ls_3}%
	\end{subfigure}
	\caption{Solutions to the cantilever beam, mid-load beam and displacement inverter mechanism problems showing mesh independence. Size of squares indicate the desired length scale}	\label{fig:Sol_MI}
\end{figure}

\begin{figure}[h!]
	\centering
	\subcaptionbox{ $ls = 1$
	}{\includegraphics[width=0.45\textwidth]{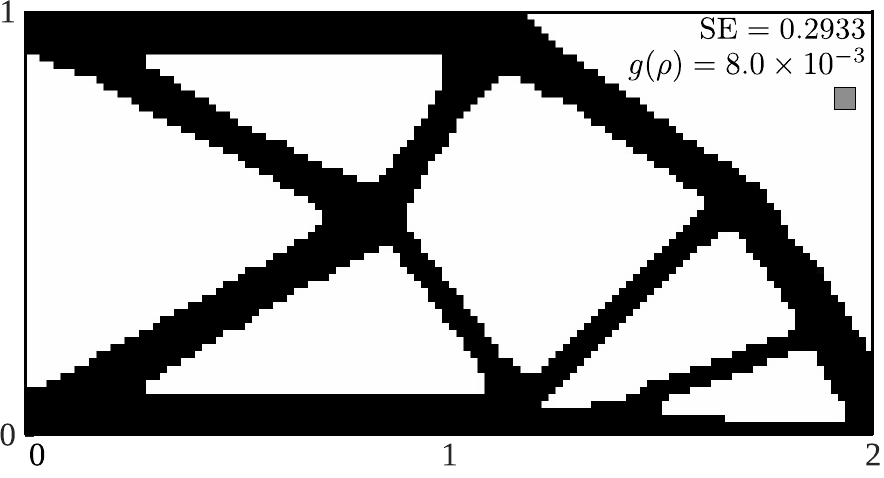}}
	\label{fig:Sol_Canti_120x60_vf_35_ls_1}
	\hspace{0.4cm}
	\subcaptionbox{ $ls = 2$
	}{\includegraphics[width=0.45\textwidth]{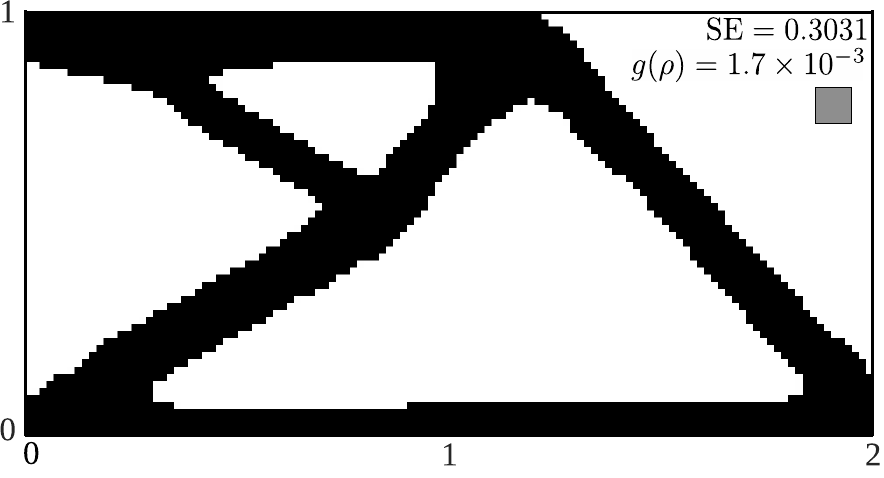}}
	\label{fig:Sol_Canti_120x60_vf_35_ls_2}%
	\\
	\subcaptionbox{ $ls = 3$
	}{\includegraphics[width=0.45\textwidth]{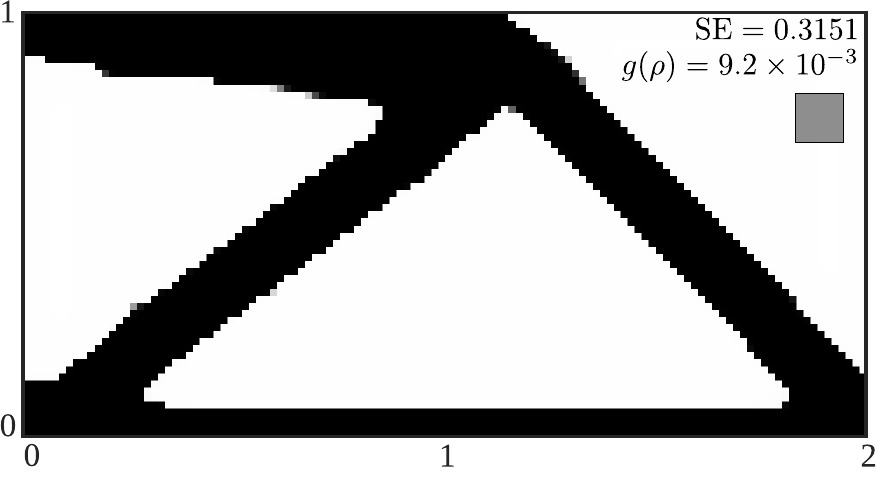}}
	\label{fig:Sol_Canti_120x60_vf_35_ls_3}%
	\hspace{0.4cm}
	\subcaptionbox{Convergence history for solution shown in Fig.~\ref{fig:Sol_Canti}a. }{\includegraphics[width=0.45\textwidth]{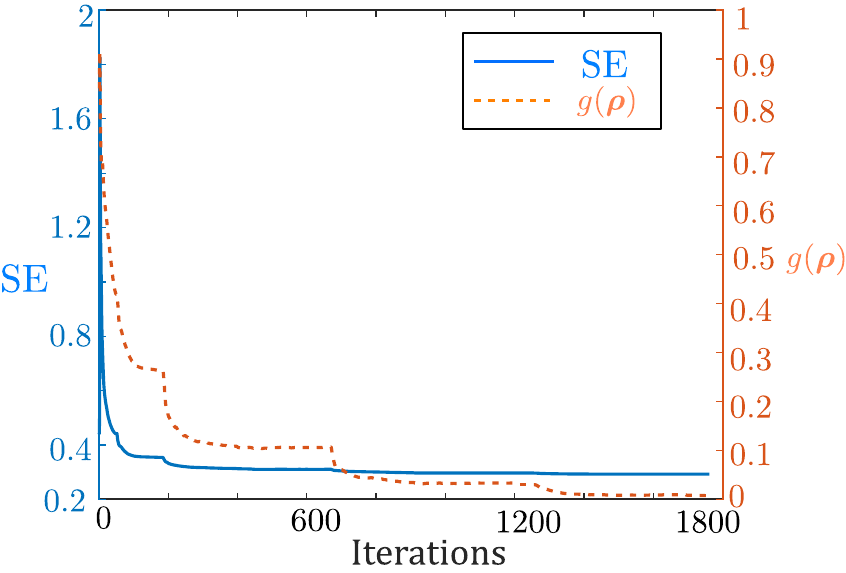}}
	\label{fig:Sol_Canti_120x60_vf_35_ls_1_convergence}
	\caption{Solutions to the cantilever beam problem. Mesh size: $120\times 60$,\, $v_f = 0.35$,\, $\mu_s = 10^3$. Size of squares indicate the desired length scale}
	\label{fig:Sol_Canti}
\end{figure}
\begin{figure}[h!]
	\centering
	\subcaptionbox{$v_f = 0.35$ }{\includegraphics[width=0.45\textwidth]{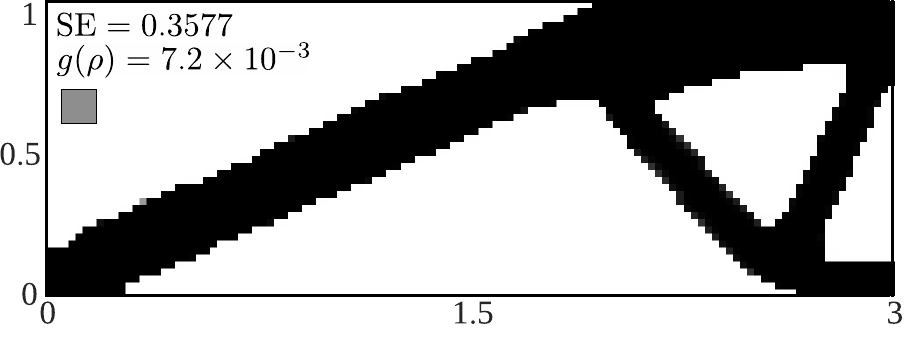}}
	\label{fig:Sol_MBB_120x40_vf_35_ls_2}%
	\hspace{0.4cm} 
	\subcaptionbox{$v_f = 0.25$ }{\includegraphics[width=0.45\textwidth]{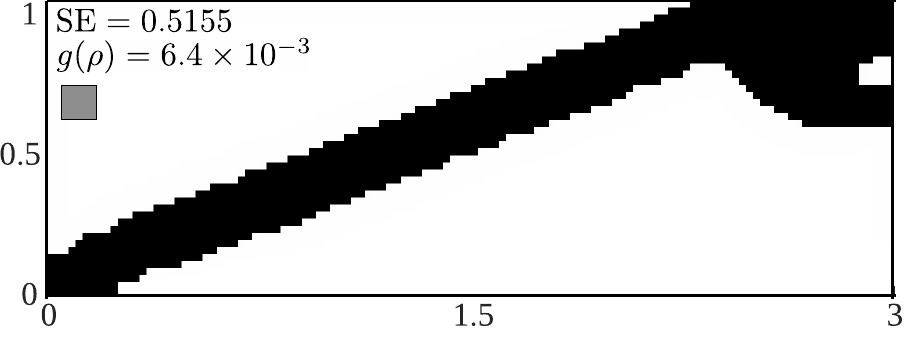}}
	\label{fig:Sol_MBB_120x40_vf_25_ls_2}%
	\\
	\subcaptionbox{ $v_f = 0.18$ }{\includegraphics[width=0.45\textwidth]{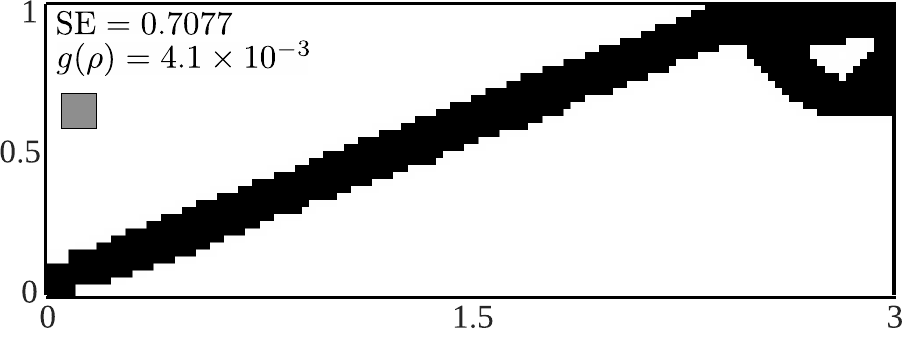}}
	\label{fig:Sol_MBB_120x40_vf_18_ls_2}
	\hspace{0.4cm}
	\subcaptionbox{Convergence history for solution displayed in Fig.~\ref{fig:Sol_MBB}a }{\includegraphics[width=0.45\textwidth]{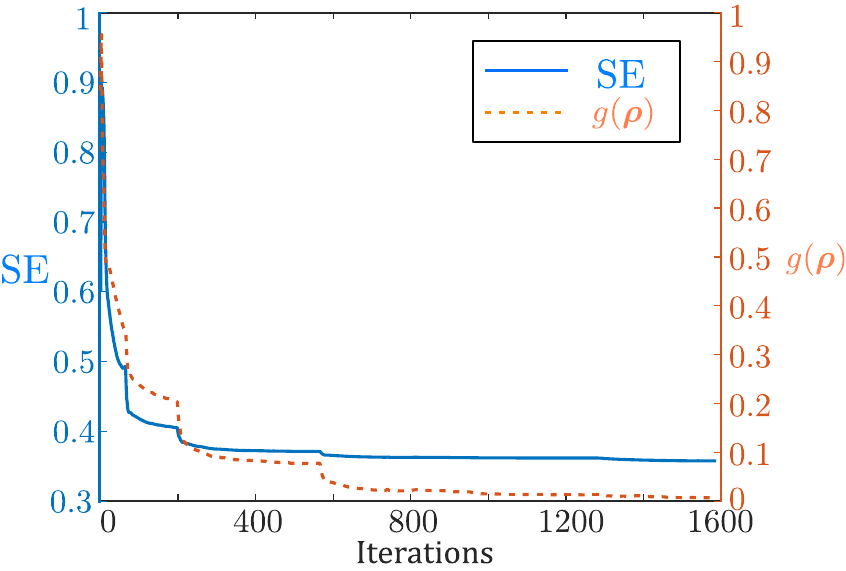}}
	\label{fig:Sol_MBB_120x40_vf_35_ls_2_convergence}
	\caption{Solutions to the mid-load beam problem.  Mesh size: $120\times 40,\,ls = 2$,\, $\mu_s =10^3$. Size of squares indicate the desired length scale.}
	\label{fig:Sol_MBB}
\end{figure}
\begin{figure}[h!]
	\centering
	\subcaptionbox{$ v_f = 0.20,\, ls = 1$}{\includegraphics[width=0.45\textwidth]{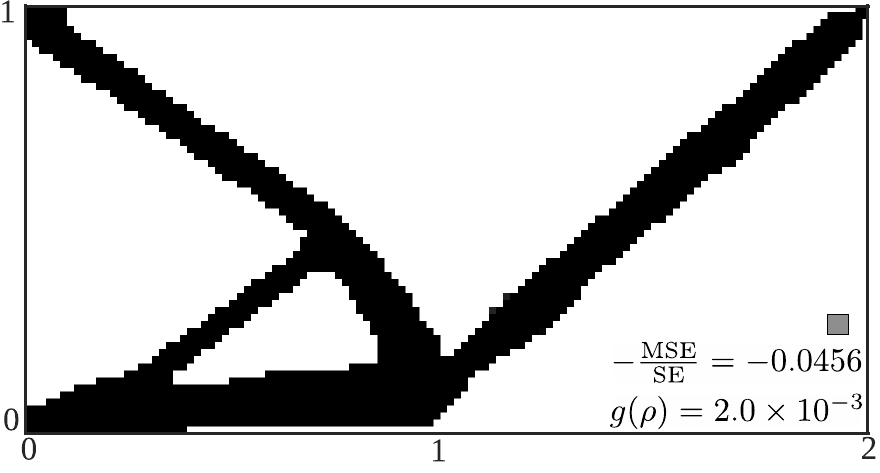}}
	\label{fig:Sol_Disp_inverter_120x60_vf_20_ls_1}%
	\hspace{0.4cm}
	\subcaptionbox{ $v_f = 0.20,\, ls = 2 $}{\includegraphics[width=0.45\textwidth]{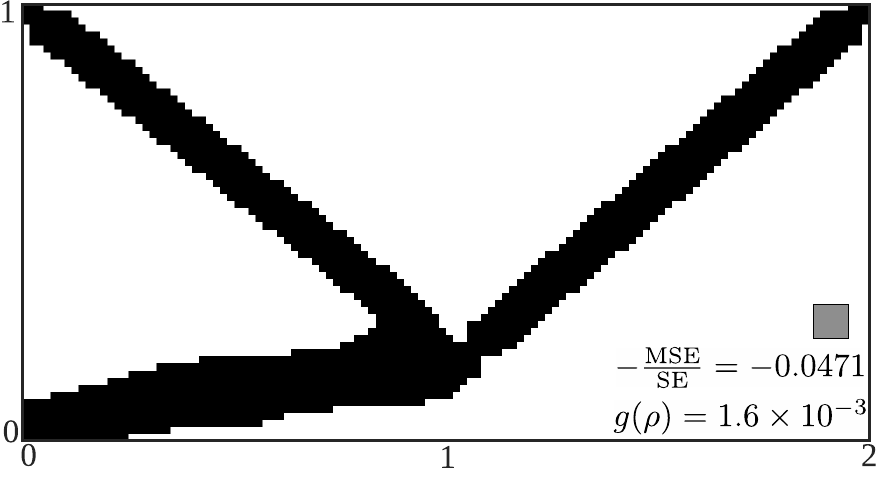}}
	\label{fig:Sol_Disp_inverter_120x60_vf_20_ls_2}
	\\
	\subcaptionbox{$ v_f = 0.30,\, ls = 2 $ }{\includegraphics[width=0.45\textwidth]{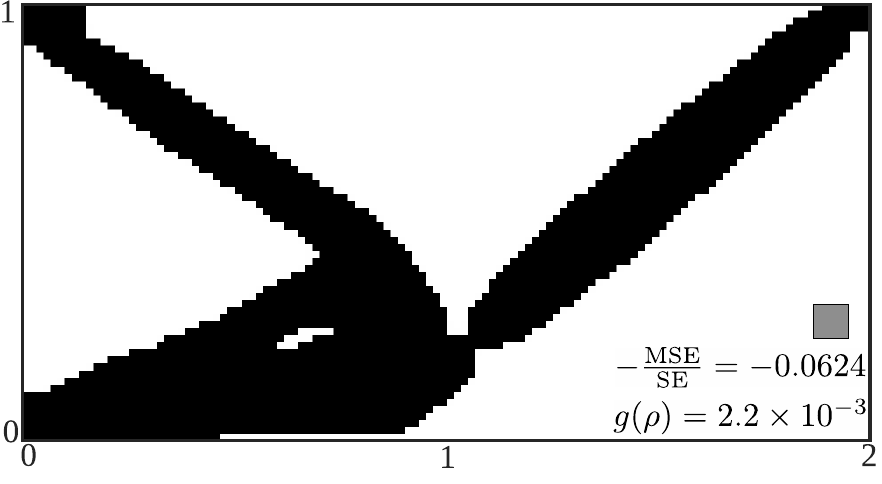}}
	\label{fig:Sol_Disp_inverter_120x60_vf_30_ls_3}
	\hspace{0.4cm}
	\subcaptionbox{Convergence history for solution depicted in Fig. \ref{fig:Sol_Disp_inverter}c }{\includegraphics[width=0.45\textwidth]{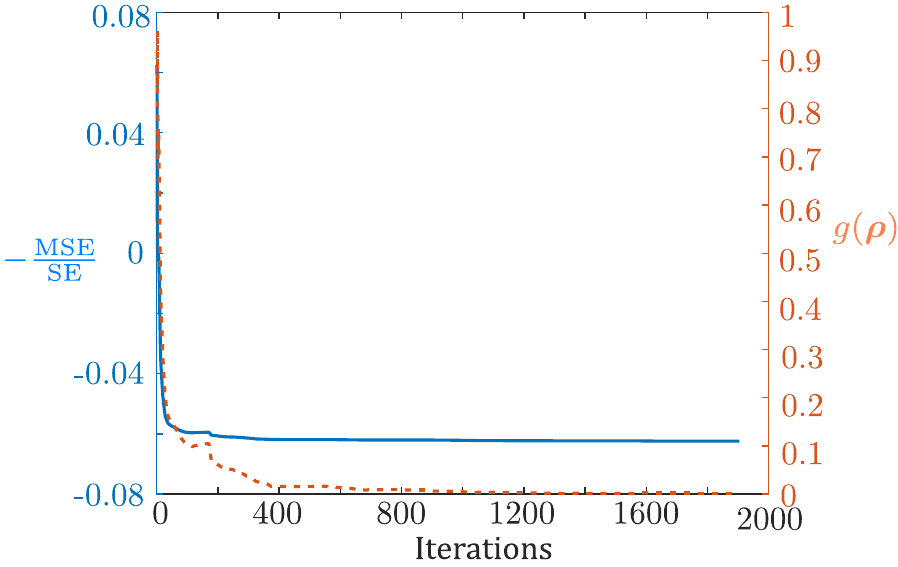}}
	\label{fig:Sol_Disp_inverter_120x60_vf_30_ls_3_converge}
	\caption{Solutions to the displacement inverter problem. Mesh size: $120\times 60,\, \mu_{CM} =10^5$. Size of squares indicate the desired length scale.}
	\label{fig:Sol_Disp_inverter}
\end{figure}
We parameterize the 2D design domains using quadrilateral elements. The square neighborhoods are selected to determine the element's density (Eq.~\ref{Eq:nFP_density_beta}) with the embedded minimum length scale. However, one may choose different neighborhood shapes~\cite{svanberg2013density}. Fig.~\ref{fig:Lenght_Neighbourhood_points} depicts the square neighborhood $\Gamma(\boldsymbol{\mathrm{X}}_i)$ on three different mesh refinements. Discretization of the neighborhood is indicated using the variable $ls$, where $ls = k$ implies $2k+1$ elements are used to discretize the edge of the square neighborhood. Fig.~\ref{fig:ls=1}, Fig.~\ref{fig:ls=2} and Fig.~\ref{fig:ls=4} depict the neighborhood discretization for $ls =1,\,2,\,\text{and}\, 4$, respectively and thereby employing $9,\,25,\,\text{and}\, 81$ elements, respectively.  
The choice of a square neighborhood has the benefit of eradicating approximation error in defining neighborhoods in opposition to the circular neighborhood when moving a problem from an analytical to a numerical setup for the quadrilateral elements.

We employ $f(\beta_j) = e^{\beta_j}$ (Table~\ref{Tab:fucntion_choices}) where $ \beta_j \leq 0$ for the density evaluation (Eq.~\ref{Eq:nFP_density_beta}) unless otherwise stated for the problems. $\beta_{lb} = -10 \times d_n$ is set,  where $d_n$ is the size of $\mathbb{N}_i$  element $i$. SIMP parameter $\eta = 3$ (Eq.~\ref{Eq:SIMP}) and minimum destiny variable $\rho_{min} = 10^{-4}$ are set (Eq.~\ref{Eq:SIMP}).  Young's modulus $E = 2\times 10^{4}$ and Poisson's ratio $\nu = 0.3$ are set. The design domains for the solved stiff structures and an inverter CM are shown in Fig.~\ref{fig:Prob_description_stiff} and Fig.~\ref{fig:Prob_description_disp_invert}, respectively. The dimensions, boundary conditions, and applied load locations for all the problems are mentioned in the figures. Because of the symmetric conditions available in the problems, we use and depict only one-half symmetric design domains for the mid-load beam (Fig.~\ref{fig:Prob_description_MLB}) and the inverter mechanism (Fig.~\ref{fig:Prob_description_disp_invert}) for the optimization purposes. The desired motion/deformation is in the opposite direction of the actuating load for the inverter mechanism.

Next, we show the mesh independence characteristic of the proposed nFP method by solving the described problems for the same volume fraction and neighborhood size at different levels of mesh refinement. The captions below each figure give mesh size and corresponding $ls$ to maintain neighborhood size. The final layout of the material distribution, along with objective values and grayness measures $g(\bm{\rho})$ for each solution, is reported.

\begin{figure}[h!]
	\centering
	\subcaptionbox{Density distribution at iteration 60}{\includegraphics[width=0.45\textwidth]{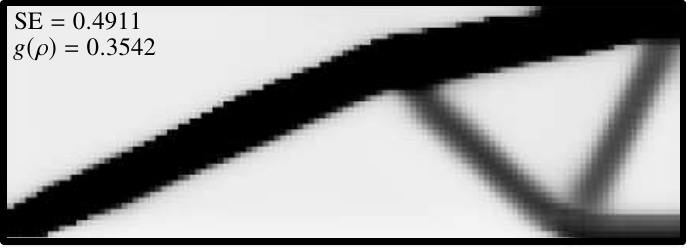}}
	\label{fig:Sol_MBB_iter_60}%
	\quad
	\subcaptionbox{Density distribution at iteration 70}{\includegraphics[width=0.45\textwidth]{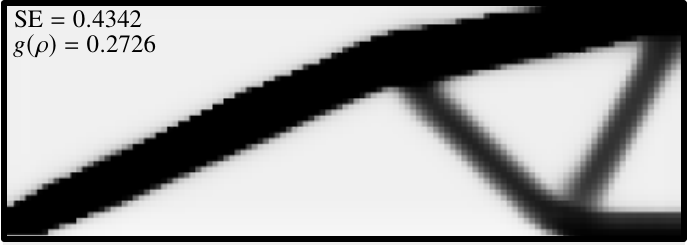}}
	\label{fig:Sol_MBB_iter_70}%
	\quad
	\subcaptionbox{Density distribution at iteration 200}{\includegraphics[width=0.45\textwidth]{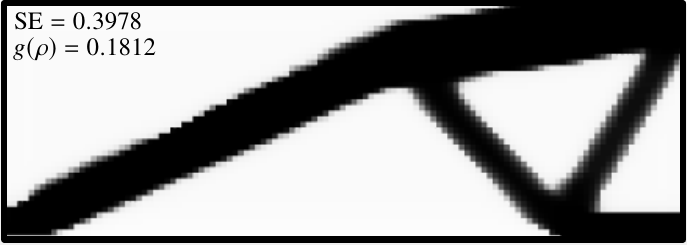}}
	\label{fig:Sol_MBB_iter_200}
	\quad
	\subcaptionbox{Density distribution at iteration 210}{\includegraphics[width=0.45\textwidth]{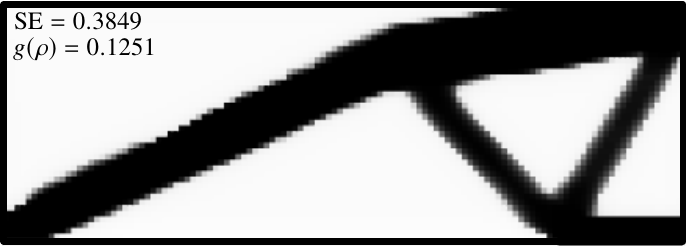}}
	\label{fig:Sol_MBB_iter_210}
	\quad
	\subcaptionbox{Density distribution at iteration 268}{\includegraphics[width=0.45\textwidth]{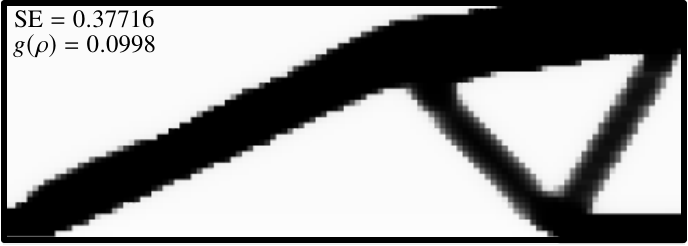}}
	\label{fig:Sol_MBB_iter_267}
	\quad
	\subcaptionbox{Density distribution at iteration 687}{\includegraphics[width=0.45\textwidth]{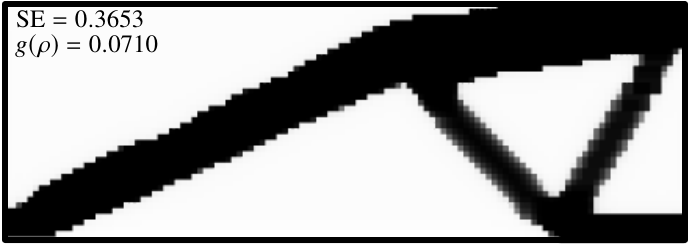}}
	\label{fig:Sol_MBB_iter_687}
	\quad
	\subcaptionbox{Density distribution at iteration 693}{\includegraphics[width=0.45\textwidth]{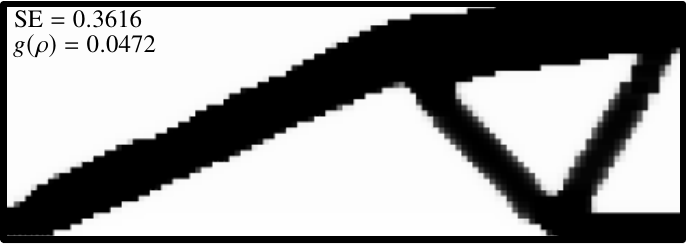}}
	\label{fig:Sol_MBB_iter_693}
	\quad
	\subcaptionbox{Density distribution at iteration 976}{\includegraphics[width=0.45\textwidth]{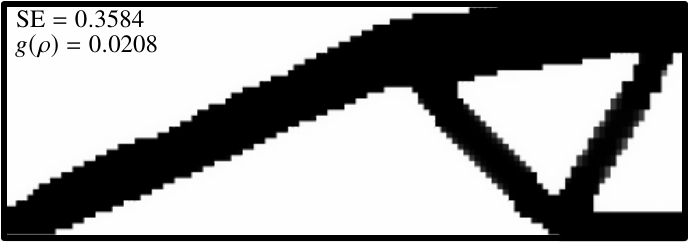}}
	\label{fig:Sol_MBB_iter_976}
	\quad
	\subcaptionbox{Density distribution at iteration 983}{\includegraphics[width=0.45\textwidth]{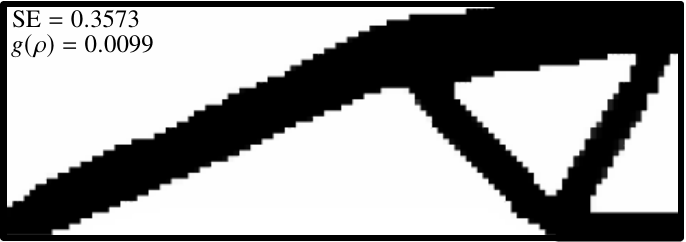}}
	\label{fig:Sol_MBB_iter_983}
	\quad
	\subcaptionbox{Density distribution at iteration 1009}{\includegraphics[width=0.45\textwidth]{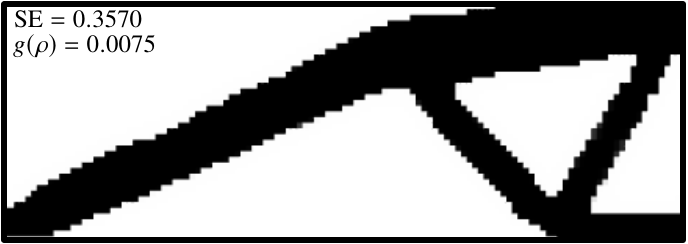}}
	\label{fig:Sol_MBB_iter_1009}
	\caption{Intermediate density distribution to the mid-load beam problem (Fig~8a).}
	\label{fig:Sol_MBB_inter}
\end{figure}
\subsection{Mesh independence}\label{Sec:MeshInd}
Mesh independence of the formulation is established by showing that identical solutions can be obtained with different mesh refinements. We select the cantilever beam (Fig.~\ref{fig:Prob_description_Canti}), mid-load beam (Fig.~\ref{fig:Prob_description_MLB}), and inverter CM (Fig.~\ref{fig:Prob_description_disp_invert}) for demonstrating the mesh independence herein.

Fig.~\ref{fig:Sol_MI} displays solutions to the cantilever beam, displacement inverter design, and mid-load beam problems at various levels of mesh refinement while retaining the neighborhood size and available volume fraction, $v_f$. The inbuilt MATLAB optimizer \texttt{fmincon} \cite{MATLAB:2010} is employed to solve each case,  with a uniform $\beta$ as the initial guess. The convergence criteria are set using \texttt{optimset} MATLAB function, wherein MaxFunEvals, i.e., maximum number of function evaluations permitted, MaxIter, i.e., the maximum number of iterations allowed, and TolFun, i.e., lower bound on the change in the value of objective, are set to 2000, 2000, and $1\times10^{-10}$, respectively.  The final topologies for the cantilever beam for the three mesh sizes are identical, and their respective final objective values are close. This consistency is observed for both the stiff structures and the compliant inverter mechanism, indicating that all solutions converge to the same or nearby continuum solution. Minor differences in shape and size could be attributed to differences in the optimizer's path to the solution. Thus, mesh independence for the formulation can be asserted, as convergence to the same local minima is achievable regardless of the mesh.

For the displacement inverter design problem, local thinning of the structure is observed. This is expected as the solution to the corresponding continuum problem is expected to have point connections. The solution in Fig. \ref{fig:Sol_MI}e appears to give a hinge, but on close inspection, a cell overlap is observed at that location, thus providing finite stiffness. The grayness values for all the cases are close to or less than 1\%, showing solutions are close to binary.
\subsection{Volume fraction and length scale}\label{Sec:VFLS}
This section provides solutions to the problems above for various volume fractions and length scales $ls$ while maintaining a consistent mesh size. The proposed nFP method applies the desired length scale (which is embedded in the density formulation) as that due to filtering~\cite{bourdin2001filters} and projection~\cite{Guest2004}.

First, the cantilever beam problem (Fig.~\ref{fig:Prob_description_Canti}) is solved with different $ls$ values while keeping the same volume fraction $v_f = 0.35$. The results are reported in Fig.~\ref{fig:Sol_Canti}. One notes that as $ls$ increases, the size of the members increases. In addition, topologies are different for $ls=1$ (Fig.~\ref{fig:Sol_Canti}a), $ls=2$ (Fig.~\ref{fig:Sol_Canti}b) and $ls=3$ (Fig.~\ref{fig:Sol_Canti}c). These observations are expected and similar to all the previously reported TO approaches. With $ls=1$, the objective value is lower than those obtained with  $ls=2$ and $ls=3$. Note that the length scale is violated at the boundary, which should be the case, as we have not employed boundary padding~\cite{clausen2017filter,kumar2021numerical}. 

Second, the mid-load problem (Fig.~\ref{fig:Prob_description_MLB}) is solved for different volume fractions while keeping $ls$ same. The results are displayed in Fig.~\ref{fig:Sol_MBB}. As expected, reducing the available volume fraction leads to fewer structural members and worse objective function values. Even for low-volume fractions, the method resulted in well-connected structures that adhere to the length scale measure provided. The grayness measure obtained is relatively low, showcasing the method's ability to work for relatively low-volume fractions.

Third, the displacement inverter compliant mechanism (Fig.~\ref{fig:Prob_description_disp_invert}) is solved for different $ls$ and volume fractions $v_f$. The results are shown in Fig.~\ref{fig:Sol_Disp_inverter}. Fig.~\ref{fig:Sol_Disp_inverter}a and Fig.~\ref{fig:Sol_Disp_inverter}b report solutions for the same volume fraction with  $ls=1$  and $ls=2$, respectively. Fig.~\ref{fig:Sol_Disp_inverter}b and Fig.~\ref{fig:Sol_Disp_inverter}c provide the effects of different volume fractions with the same length scale $ls=2$. The above solutions indicate the success of the proposed nFP method for different volume fractions with various length scales. Next, we mention the convergence and grayness measure characteristics of the nFP approach.

\subsubsection{Convergence and grayness measure}\label{Sec:cg}
Figures~\ref{fig:Sol_Canti}d, \ref{fig:Sol_MBB}d and \ref{fig:Sol_Disp_inverter}d display the convergence history of the solutions reported in Figs. \ref{fig:Sol_Canti}a, \ref{fig:Sol_MBB}a and \ref{fig:Sol_Disp_inverter}c, respectively. One notes that the objective value converges much earlier than the solutions' grayness, $g(\rho)$. Some sudden jumps are observed in the objective and grayness values. These jumps occur at almost the same time for both functions and are usually associated with the formation of structural members. To visualize this, we present the intermediate solutions for the mid-load beam problem (Fig.~\ref{fig:Prob_description_MLB}) in Fig.~\ref{fig:Sol_MBB_inter} for iterations 60, 70, 200 and 210. A sudden drop in objective and grayness for the problem are observed close to iteration 60 (Fig.~\ref{fig:Sol_MBB}d) and 200 (Fig.~\ref{fig:Sol_MBB}d). Both objective and grayness measures for each of the intermediate solutions are mentioned in the figure (Fig.~\ref{fig:Sol_MBB}), along with the density distribution. A significant drop in both values is observed between iterations 60 and 70 and 200 and 210. Intermediate solutions reveal that sudden dips in objective value are associated with the reinforcement of slant edges in the structure during topology optimization.

One notes that the optimization processes for these problems converge close to 200 iterations given objective values; however, their grayness measure is still on a decreasing path. We further keep running the optimization to learn how long optimization can take to achieve lower $g(\bm{\rho})$ values without reinforcing. As optimization progresses, $\beta_j$ increases, and the magnitude of the gradients decreases; thus, convergence becomes slow. The convergence of the grayness measure also depends upon the user-selected $f(\beta_j)$ (see Sec.~\ref{Sec:FunctionC}).   

Typically, a continuation scheme, a heuristic-based approach, is employed in TO to achieve \textit{high-quality} local minima solutions characterized by low $g(\bm{\rho})(\approx0.1\%)$ with good convergence~\cite{Guest2004,sigmund_maute_2013}. However, the proposed nFP achieves high-quality local minima solutions without using any continuation schemes, as evident in Figs.~\ref{fig:Sol_MI}--\ref{fig:Sol_Disp_inverter}, \ref{fig:Sol_func_choices}, \ref{fig:MBB_sol_3d}--\ref{fig:Sol_nFP_TKD}. This aspect is also supported by the intermediate results depicted in Fig.~\ref{fig:Sol_MBB_inter}. At $60^\text{th}$ iteration, a low-quality ($g(\bm{\rho})$ =35.42\%) solution is obtained. Nevertheless, as optimization progresses, the solution tends towards a high-quality solution (Fig.~\ref{fig:Sol_MBB_inter}). The numerical results indicate that allowing for the natural attainment of solutions close to 0-1 requires over 1000 iterations (Figs.~\ref{fig:Sol_MI}--\ref{fig:Sol_Disp_inverter}). However, it is possible to terminate the optimization process using tolerances on $g(\bm{\rho})$, such as $g_{tol}=0.1$ or low, requiring less number of iteration for high $g_{tol}$. With  $g_{tol}=0.1$, the optimization gets terminated at $268$ (Fig.~\ref{fig:Sol_MBB_inter}f), i.e., a solution with $g(\bm{\rho}) = 9.979\%$.  Results in Fig.~\ref{fig:Sol_MBB_inter}f ($687^\text{th}$ iteration), Fig.~\ref{fig:Sol_MBB_inter}g ($693^\text{th}$ iteration), Fig.~\ref{fig:Sol_MBB_inter}h ($976^\text{th}$ iteration), Fig.~\ref{fig:Sol_MBB_inter}i ($983^\text{th}$ iteration), and Fig.~\ref{fig:Sol_MBB_inter}j ($1009^\text{th}$ iteration) are displayed for $g_{tol}$ equals to  0.075, 0.05, 0.025, 0.001 and 0.0075, respectively. One notices that lower $g_{tol}$ requires more iterations because, as optimization progresses,  $\beta_j$ increases (an elemental density variable--local quantity, cf. Eq.~\ref{Eq:nFP_density_beta}), i.e., magnitude of the gradient of the objective diminish (this observation is also true for the method presented in~\cite{Guest2004} (see Appendix.~\ref{App:A1}), where $\beta$ parameter of the method is a global variable.); thus, convergence becomes slow. The above discussion indicates that the method avoids low-quality local minima solutions without any continuation scheme.

The grayness measures for the presented solutions are close to or less than $1\%$. Almost all solutions exhibit some gray elements at the solid-void interface. From convergence histories, it is realized that the decline in $g(\rho)$ is consistent but gradual, and therefore, eliminating all gray elements will take a significantly large number of iterations. Some gray elements remain in the final solution due to low magnitudes of gradients, and computational local minima are achieved while gray cells exist within the density distribution.

For the most part, the solid-void interface lacks transition regions; elements with intermediate densities do not separate elements with densities close to 1 and 0. This affirms that the nFP method can produce transition-free interface solutions without requiring parameters, continuation, or user intervention. 

The grayness measure exhibits a monotonic, gradual decline with each iteration, even though no specific actions are performed to reduce the grayness measure apart from utilizing the SIMP material model, affirming that the solution naturally gravitates towards binary topologies. This demonstrates the nFP method's capability to yield close to binary solutions automatically without requiring any density threshold/parameter to enforce the optimization towards the binary solutions.

\begin{figure}[h]
	\centering
	\subcaptionbox{$f(\beta_j) = \dfrac{1}{\beta_j^n}$ }{\includegraphics[width=0.32\textwidth]{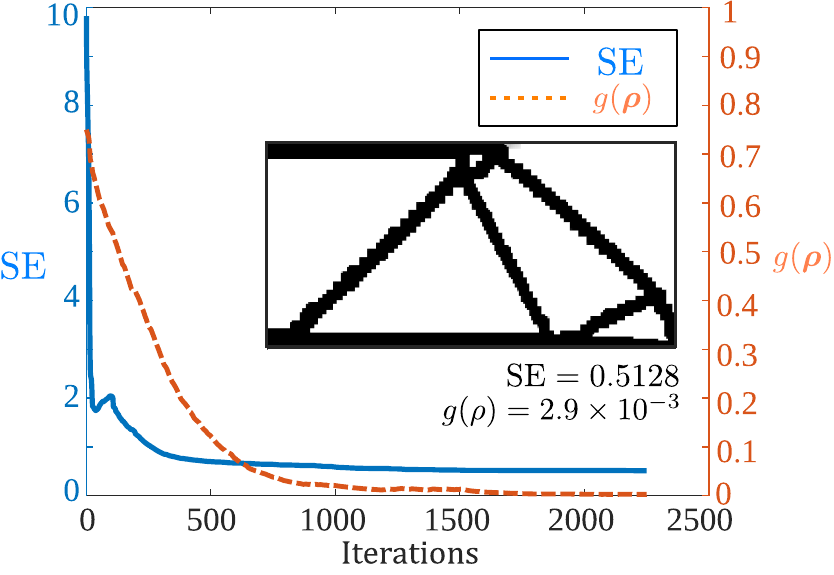}}
	\label{fig:Sol_Canti_frac_exp_200x100_vf_25_ls_2}%
	\subcaptionbox{$f(\beta_j) = 1 - \dfrac{\tan^{-1} \beta_j}{\pi/2}$}{\includegraphics[width=0.32\textwidth]{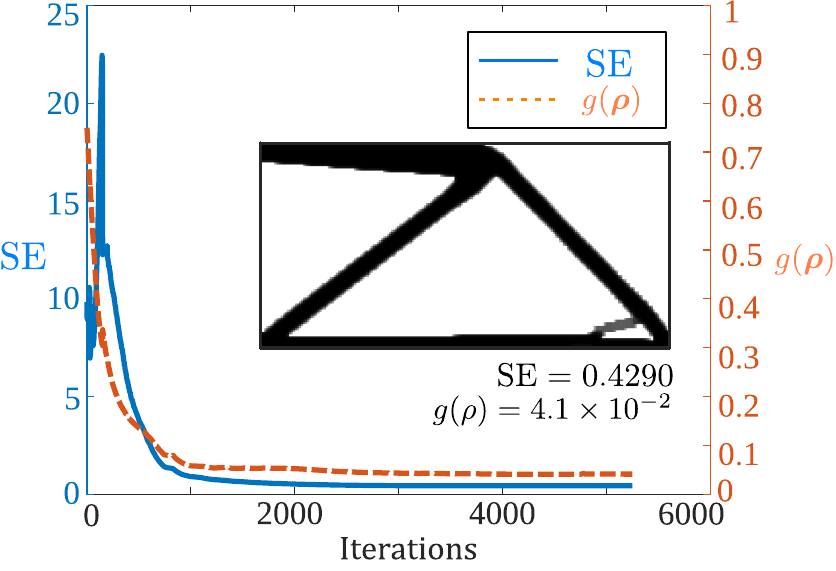}}
	\label{fig:Sol_Canti_tan_inv_200x100_vf_25_ls_2}%
	\subcaptionbox{$f(\beta_j) = 1 - \tanh \beta_j$ }{\includegraphics[width=0.32\textwidth]{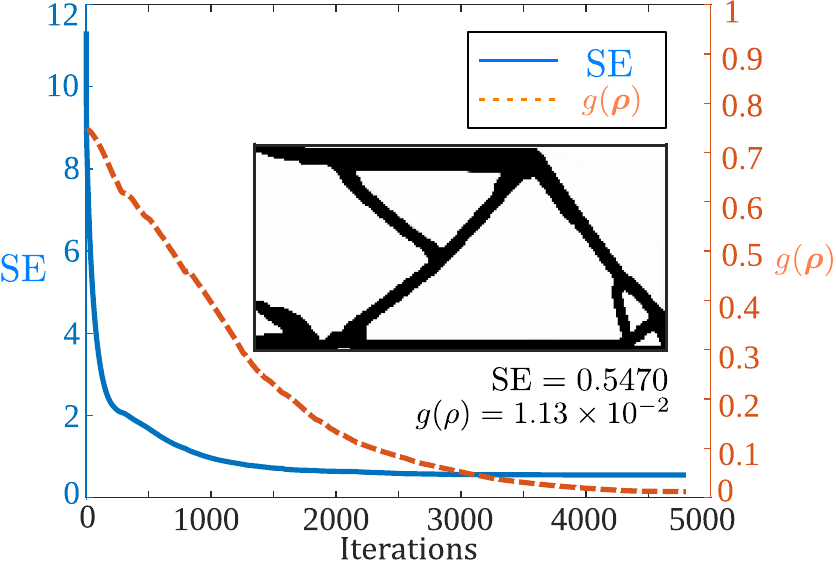}}
	\\
	\subcaptionbox{$f(\beta_j) = \dfrac{1}{\beta_j^n}$}{\includegraphics[width=0.32\textwidth]{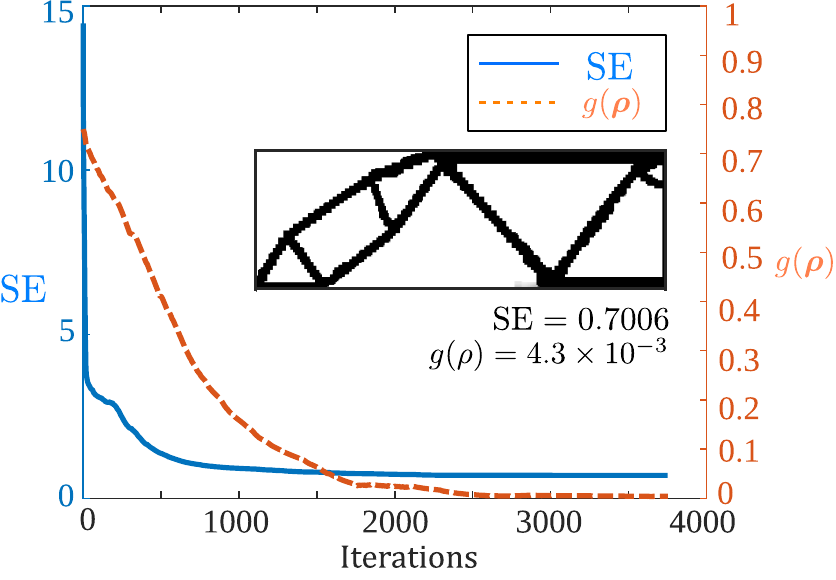}}
	\label{fig:Sol_MLB_frac_exp_300x100_vf_25_ls_2}%
	\subcaptionbox{$f(\beta_j) = 1 - \dfrac{\tan^{-1} \beta_j}{\pi/2}$ }{\includegraphics[width=0.32\textwidth]{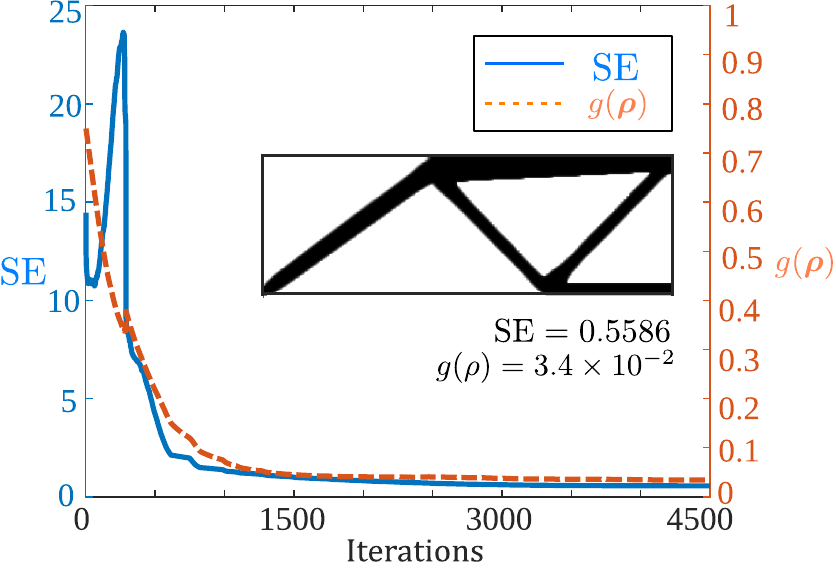}}
	\label{fig:Sol_MLB_tan_inv_300x100_vf_25_ls_2}%
	\subcaptionbox{$f(\beta_j) = 1 - \tanh \beta_j$}{\includegraphics[width=0.32\textwidth]{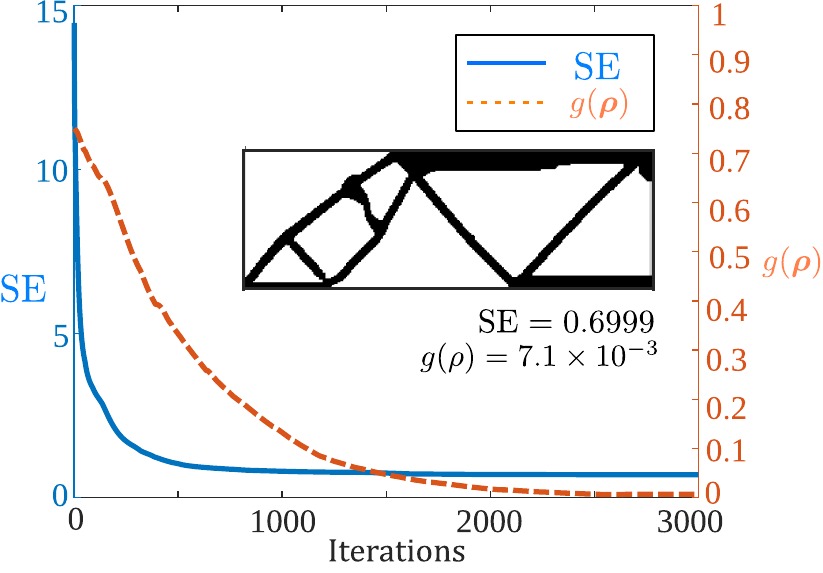}}
	\\
	\subcaptionbox{$f(\beta_j) = \dfrac{1}{\beta_j^n}$}{\includegraphics[width=0.32\textwidth]{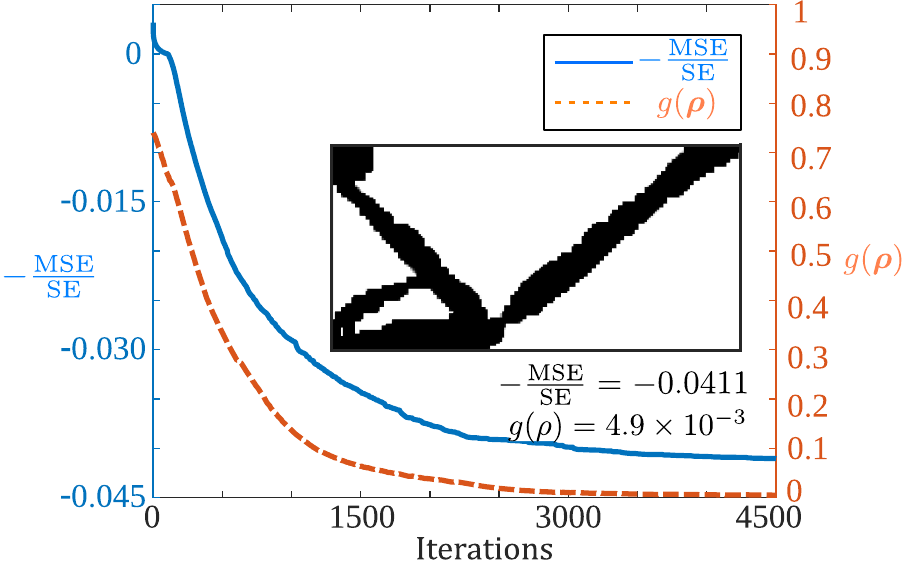}}
	\label{fig:Sol_Disp_invertor_frac_exp_200x100_vf_25_ls_2}%
	\subcaptionbox{$f(\beta_j) = 1 - \dfrac{\tan^{-1} \beta_j}{\pi/2}$ }{\includegraphics[width=0.32\textwidth]{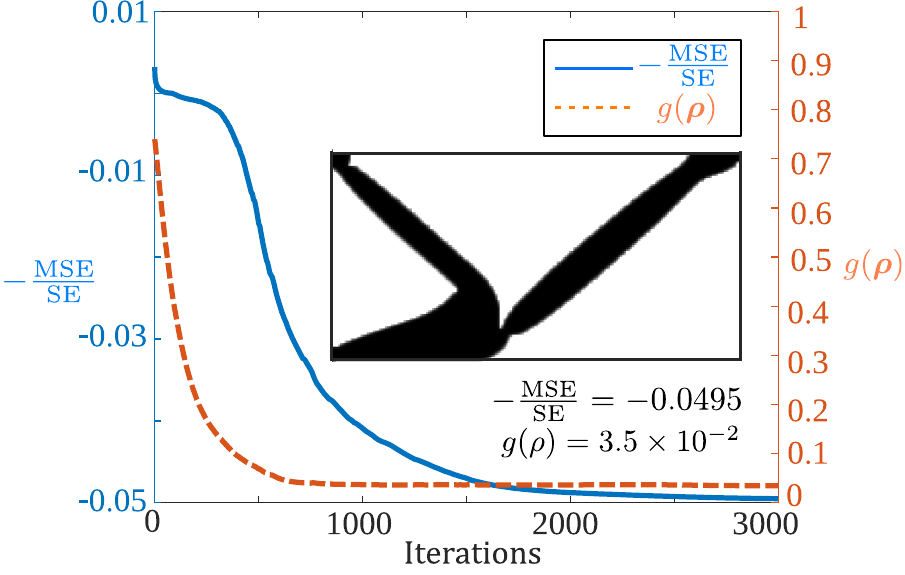}}
	\label{fig:Sol_Disp_invertor_tan_inv_200x100_vf_25_ls_2}%
	\subcaptionbox{$f(\beta_j) = 1 - \tanh \beta_j$}{\includegraphics[width=0.32\textwidth]{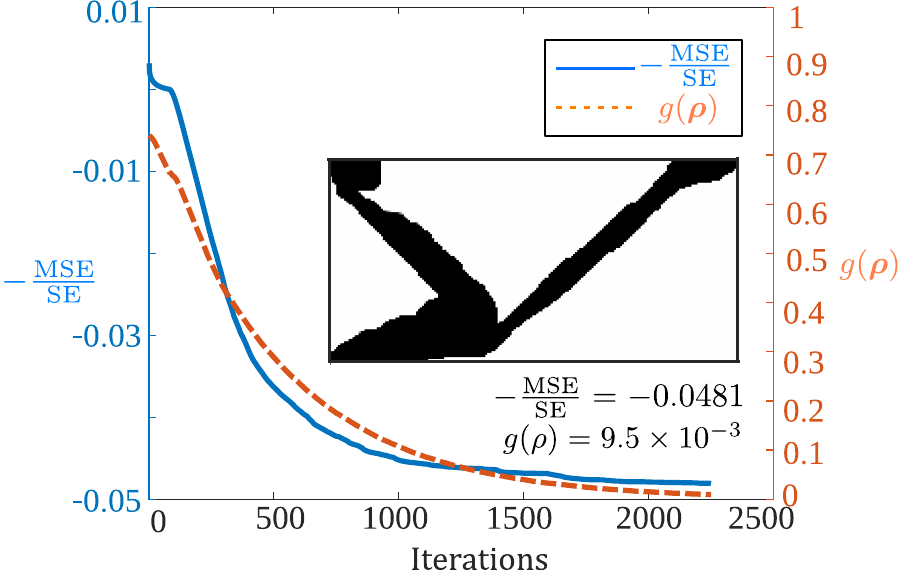}}
	\label{fig:Sol_Disp_invertor_tanh_200x100_vf_25_ls_2}
	\caption{Solutions to cantilever beam, mid-load beam, and displacement inverter mechanism are presented in the first, second, and third rows with different $f(\beta)_j$. Each column presents solutions for a specific $f(\beta_j)$. $ 200 \times 100$ elements parameterize the design domains of the cantilever beam and displacement inverter mechanism, while $300 \times 100$ elements are used for the mid-load problem. Volume fraction, $vf = 0.25$, and length scale, $ls =2$ are set. $S = 0.002$ is used (Eq.~\ref{Eq:MMAapplication})}.
	\label{fig:Sol_func_choices}
\end{figure}
\begin{figure}[h!]
	\centering
	\begin{tabular}{ccc}
		\includegraphics[width=0.32\textwidth]{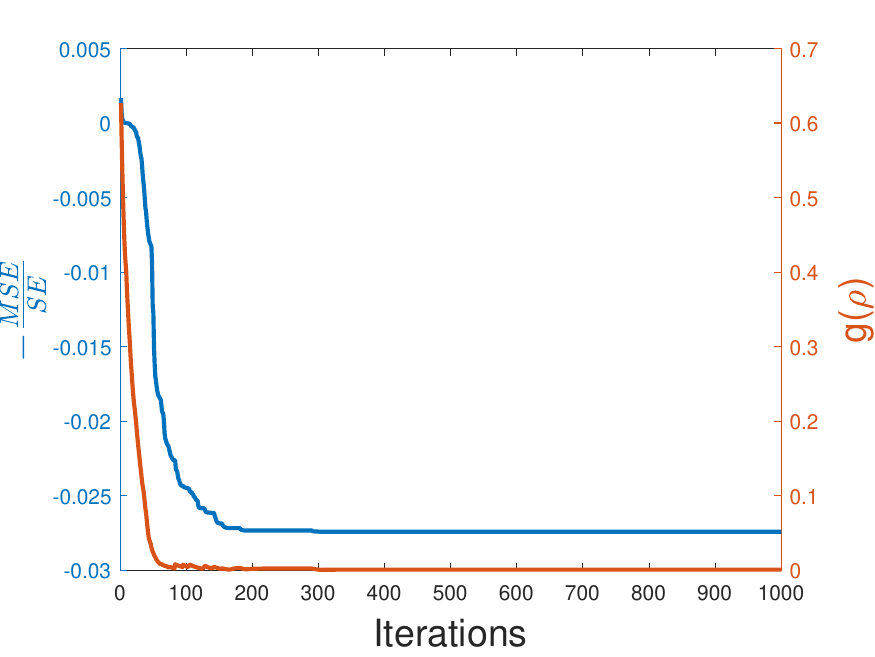}&
		\includegraphics[width=0.32\textwidth]{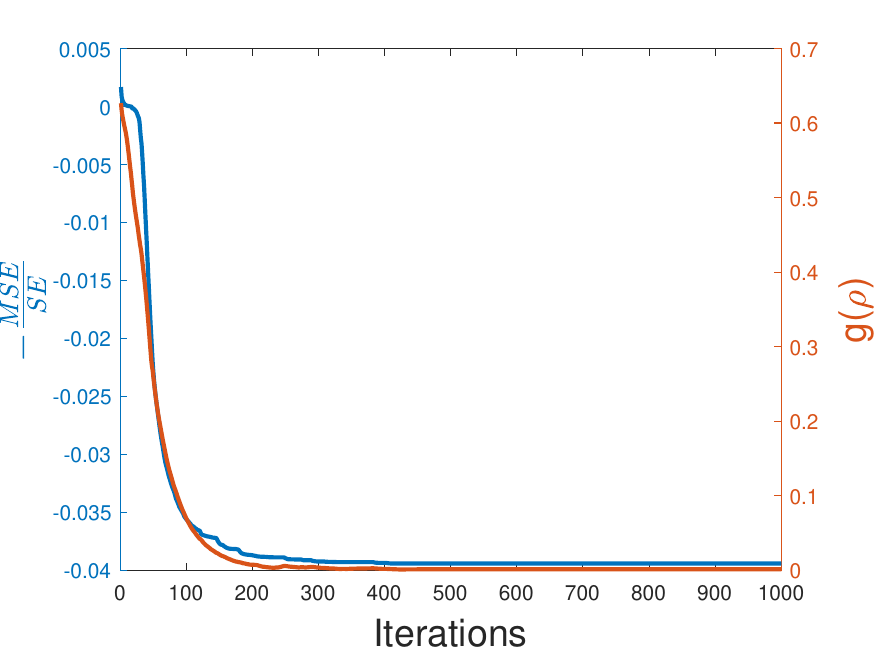}&
		\includegraphics[width=0.32\textwidth]{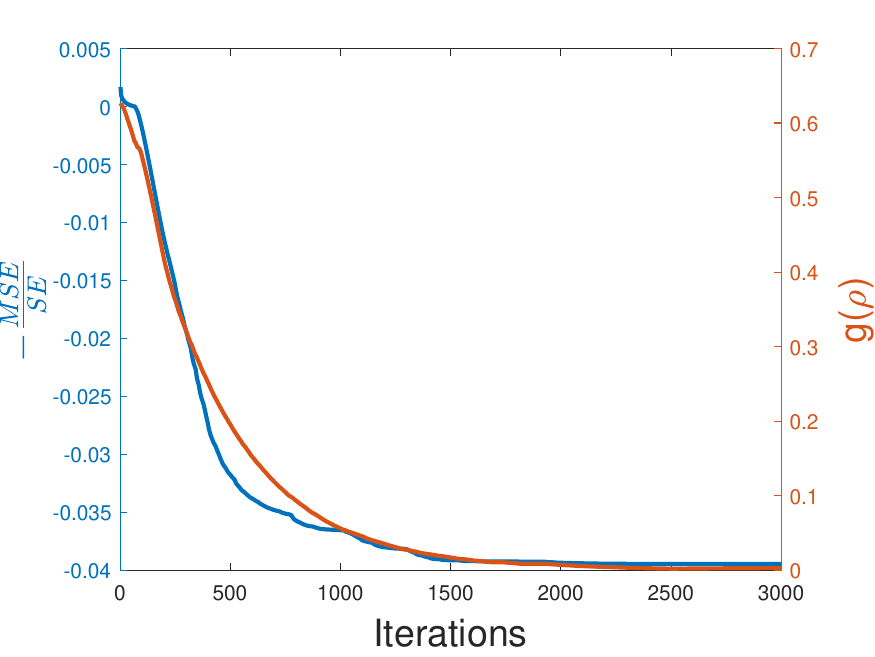}\\
		\includegraphics[width=0.28\textwidth]{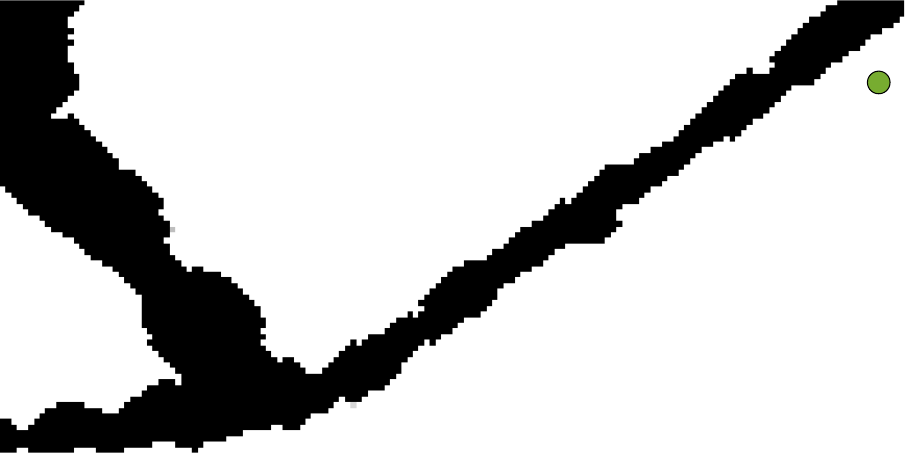}&
		\includegraphics[width=0.28\textwidth]{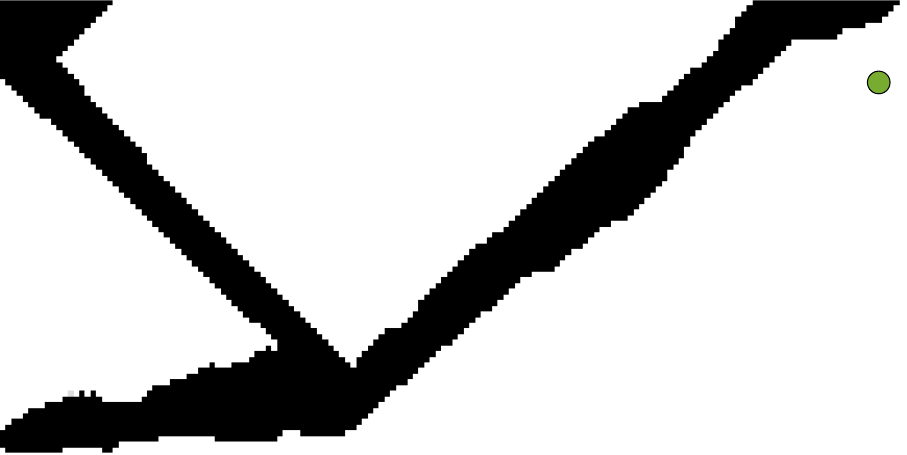}&
		\includegraphics[width=0.28\textwidth]{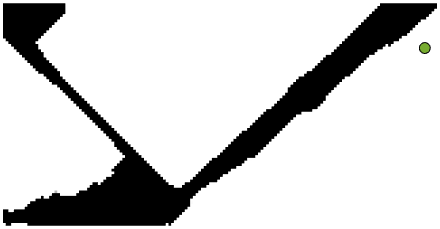}\\
		Step-size: 0.1 & Step-size: 0.025 & Step-size: 0.0025 \\
		$\frac{MSE}{SE}$ = -0.27 & $\frac{MSE}{SE}$ = -0.039 & $\frac{MSE}{SE}$ = -0.04 \\
	\end{tabular}
	\caption{Solutions to the inverter mechanism obtained with different step sizes. Top row: solution histories. Bottom Row: Inverter topologies. Size of circles indicate the desired length scale.}
	\label{fig:inv_solns_circ_DSS} 
\end{figure}
\begin{figure}[h!]
	\centering
	\subcaptionbox{}{\includegraphics[width=0.32\textwidth]{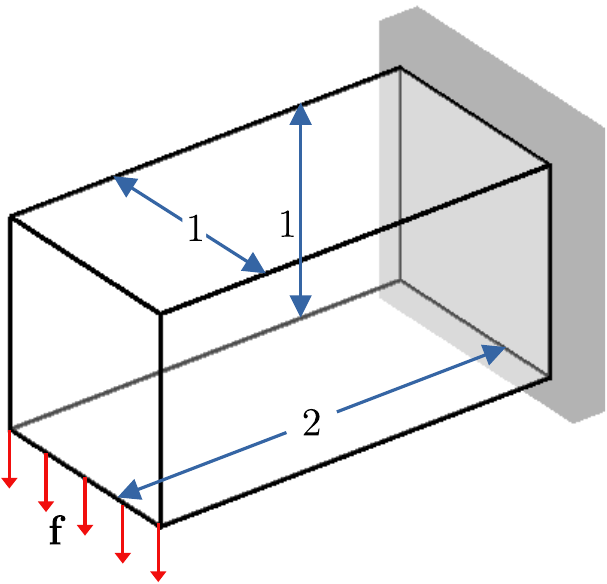}}
	\hspace{1cm}
	\subcaptionbox{}{\includegraphics[width=0.54\textwidth]{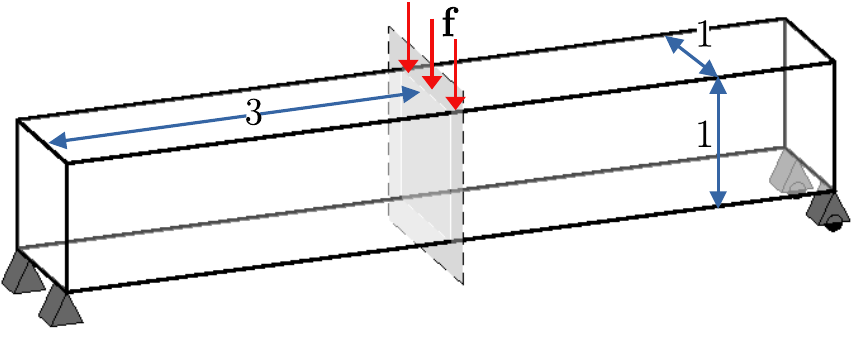}}
	\\
	\subcaptionbox{}{\includegraphics[width=0.32\textwidth]{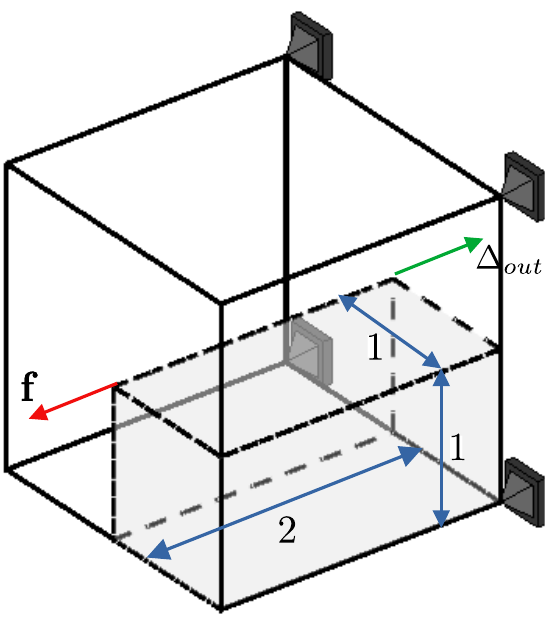}}
	\caption{Problem description for 3-dimensional (a) Cantilever beam, (b) MBB beam, and (c) Displacement inverter design problem.}
	\label{fig:Prob_description_3d}
\end{figure}
The above-mentioned numerical results support the idea that by enabling close-to-binary topologies, along with material models penalizing intermediate densities, without requiring user intervention, close-to-binary solutions can be achieved. Next, we note the solutions with different function choices (Table~\ref{Tab:fucntion_choices}).

\subsection{Function choice}\label{Sec:FunctionC}
In this section, we demonstrate the effects of the selection of different $f(\beta_j)$ in Eq.~\ref{Eq:nFP_density_beta} on the optimized designs for the problems definition presented in Fig.~\ref{fig:Prob_description_stiff} and Fig.~\ref{fig:Prob_description_disp_invert}. $ 200\times100$ elements discretize the design domains for the cantilever and displacement inverter mechanism, whereas, for the mid-load beam, $300\times 100$ elements are used. $v_f = 0.25$ and $ls = 2$ are set for all the problems. The MMA optimizer\footnote{\textit{fmincon} is not suitable for a large number of design variables because of high memory requirements~\cite{MATLAB:2010}.} is employed~\cite{svanberg1987method}. The implementation of the MMA herein is similar to the standard, except that after each MMA iteration, the new design vector is determined as~\cite{kumar2022improved}
\begin{equation}\label{Eq:MMAapplication}
	\bm{\beta}_{n} = \bm{\beta}_{o} + S (\bm{\beta}_{c} - \bm{\beta}_{o}),
\end{equation}
where $	\bm{\beta}_{n}$, $\bm{\beta}_{o}$ and $\bm{\beta}_{c}$ represent the new, old, and current design variable vectors, respectively.  $\bm{\beta}_{c}$ is the current solution provided by the MMA using $\bm{\beta}_{o}$. $S$ is the step multiplication parameter. In our experience,   $S\in[0.002,\,0.005]$ works well for the nFP settings.  The MMA optimization is terminated using the MaxIter parameter, which is chosen to provide the optimizer with a sufficient number of iterations to converge the solutions with close to $g(\bm{\rho})\approx 0.1\%$.

Figure~\ref{fig:Sol_func_choices} depicts the final solutions, objective value, grayness measure and convergence history for the cantilever beam (Fig.~\ref{fig:Prob_description_stiff}a), mid-load beam (Fig.~\ref{fig:Prob_description_stiff}b) and displacement inverter mechanism (Fig.~\ref{fig:Prob_description_stiff}c) problems for function choices $f(\beta_j) = 1 - \tanh \beta_j;~ \beta_j \in [0,\infty)$, $f(\beta_j) = \dfrac{1}{\beta_j^n};~ \beta_j \in [1,\infty),$ with $ n = 12$ and $f(\beta_j) = 1 - \dfrac{\tan^{-1} \beta_j}{\pi/2};~ \beta_j \in [0,\infty)$. 

Solutions in Fig.~\ref{fig:Sol_func_choices} demonstrate that the method yields reasonable results with different functions. 
Grayness measure for solutions generated using $f(\beta_j) = 1 - \tanh \beta_j$ and $f(\beta_j) = \dfrac{1}{\beta_j^n}$ for $n=12$, are observed to be below the desired $1\%$ while the same in not true for $f(\beta_j) = 1 - \dfrac{\tan^{-1} \beta_j}{\pi/2}$. This may be an example of computational local minima being achieved while gray elements remain within the density distribution even with high optimization iterations. This is due to the impact of the selected function on the gradient magnitude (Eq.~\ref{Eq:density_grad}). From Table~\ref{Tab:fucntion_choices}, for $f(\beta_j) = 1 - \dfrac{\tan^{-1} \beta_j}{\pi/2}$, gradient magnitudes are inversely propositional to $\beta_j$, i.e, they become low with high $\beta$. On the other hand, high $\beta$ steers the optimization towards 0-1. Thus, $f(\beta_j) = 1 - \dfrac{\tan^{-1} \beta_j}{\pi/2}$, the magnitude of gradients rapidly diminishes that, jeopardizing the optimization progress. The same observation is not true for $f(\beta_j) = 1 - \tanh \beta_j$, wherein the contribution of $f(\beta_j)$ to density gradient is bounded between $0$ and $-2$. Consequently, the function choice of $f(\beta_j) = 1 - \dfrac{\tan^{-1} \beta_j}{\pi/2}$ leads to relatively gray solutions ( Fig. \ref{fig:Sol_func_choices}b, e, and h);   the numerical optimizer finds difficult to go below a certain grayness threshold due to very low gradient magnitudes.
\begin{figure}[h!]
	\centering
	\subcaptionbox{Final density distribution}{\includegraphics[width=0.36\textwidth]{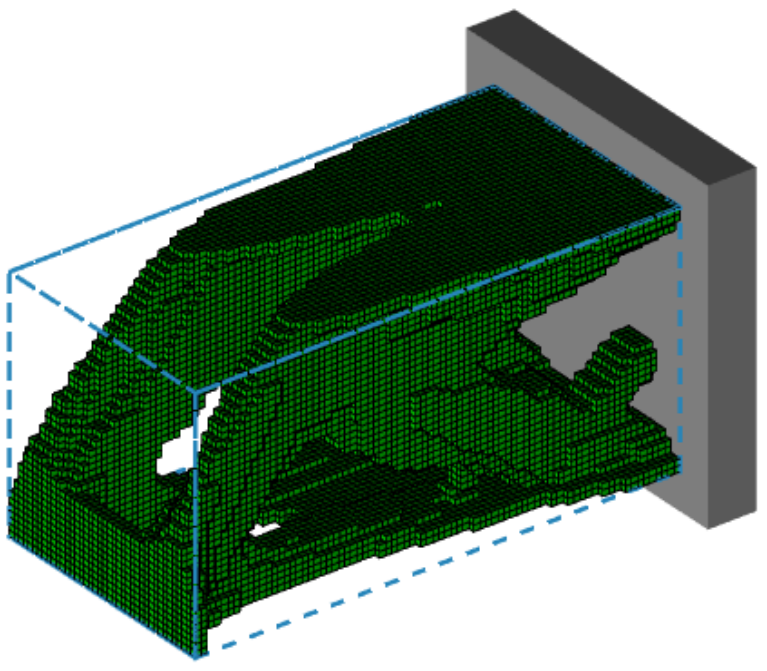}}
	\hspace{1.5cm}
	\subcaptionbox{Convergence history}{\includegraphics[width=0.4\textwidth]{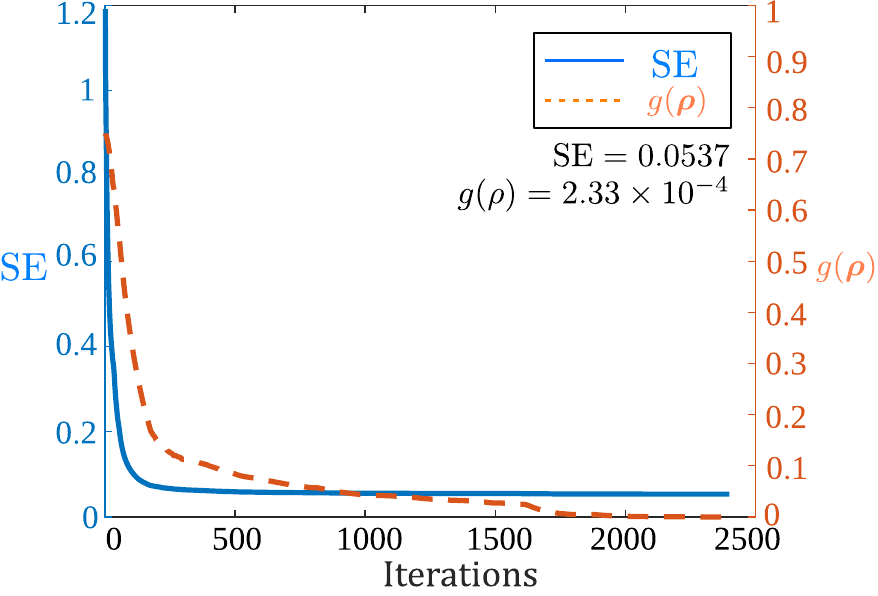}}
	\caption{Solution to the 3-dimensional cantilever beam problem using $80 \times 40 \times 40$ mesh for a volume fraction, $vf = 0.25$. MaxIter = 2500.}
	\label{fig:Canti_sol_3d}
\end{figure}

\begin{figure}[h!]
	\centering
	\subcaptionbox{Final density distribution}{\includegraphics[width=0.42\textwidth]{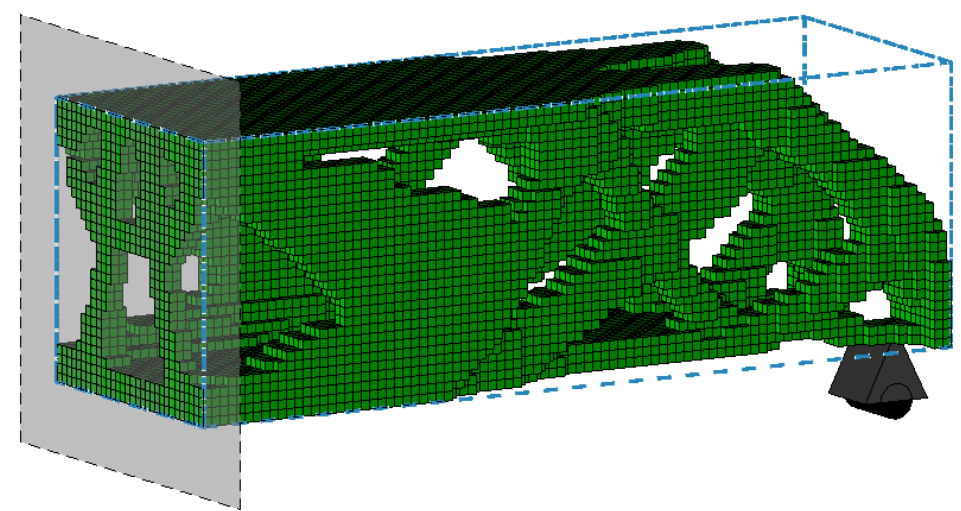}}
	\hspace{0.5cm}
	\subcaptionbox{Final structure over the complete domain.}{\includegraphics[width=0.45\textwidth]{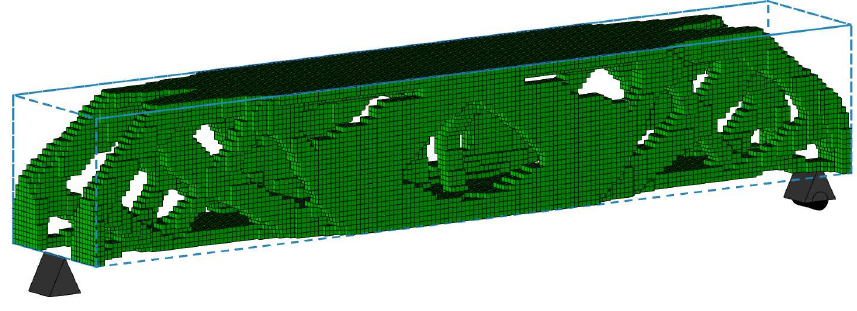}}
	\\
	\subcaptionbox{Convergence history}{\includegraphics[width=0.5\textwidth]{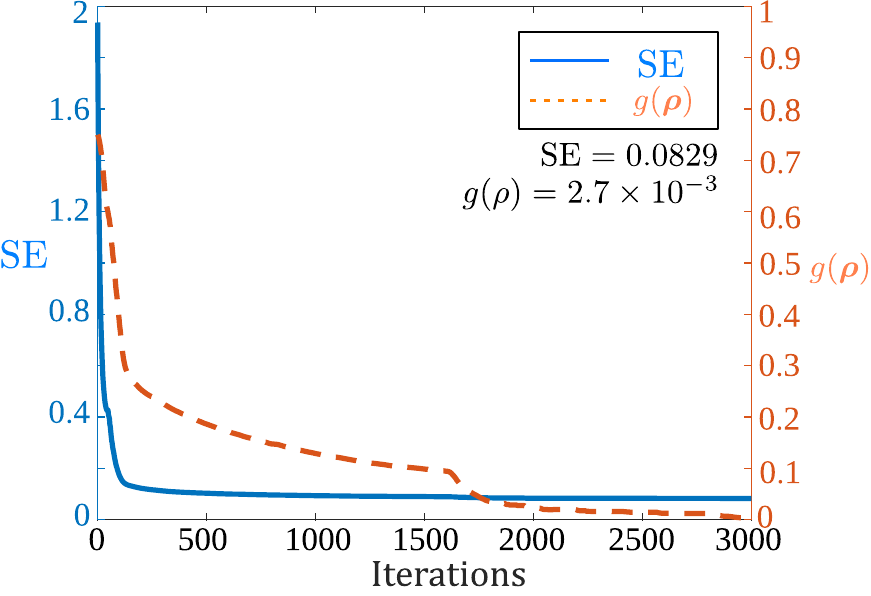}}
	\caption{Solution for the 3-dimensional MBB beam problem using $90 \times 30 \times 30$ mesh for a volume fraction, $vf = 0.25$. MaxIter = 3000.}
	\label{fig:MBB_sol_3d}
\end{figure}

\begin{figure}[h!]
	\centering
	\subcaptionbox{Final density distribution}{\includegraphics[width=0.4\textwidth]{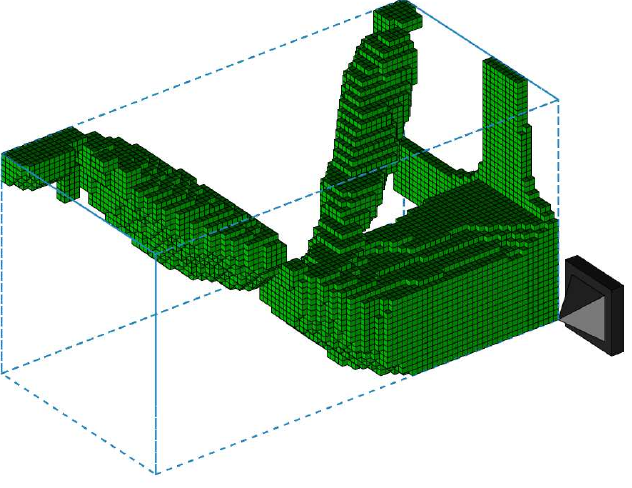}}
	\hspace{0.5cm}
	\subcaptionbox{Final structure over the complete domain.}{\includegraphics[width=0.45\textwidth]{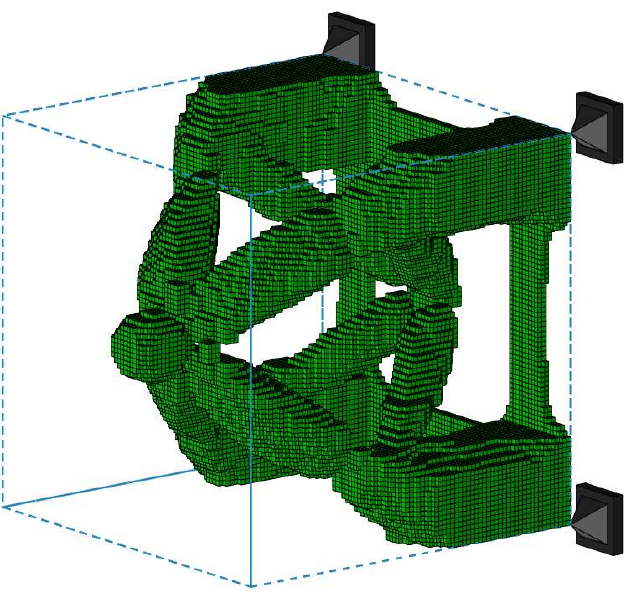}}
	\\
	\subcaptionbox{Convergence history}{\includegraphics[width=0.65\textwidth]{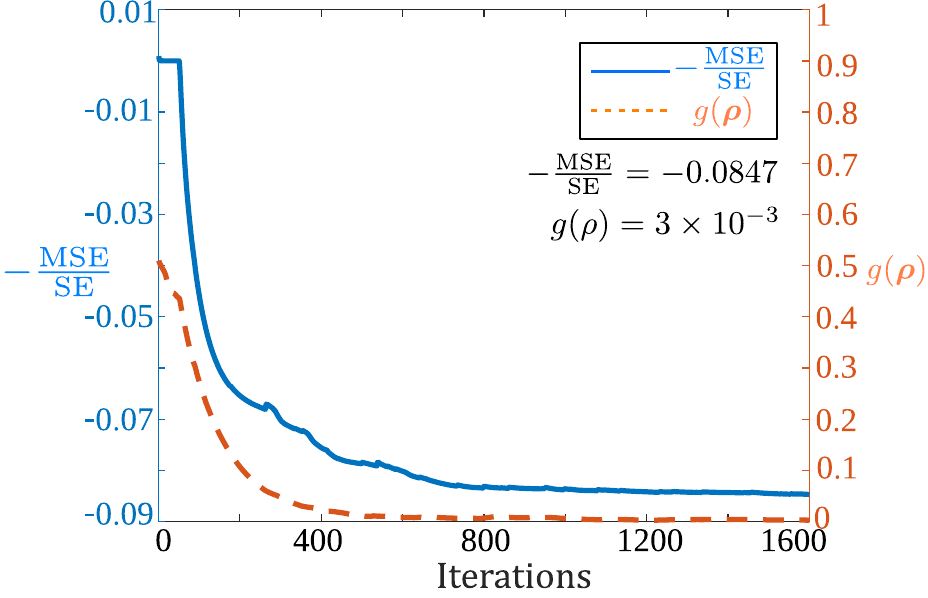}}
	\caption{Solution to the displacement inverter design problem using $80 \times 40 \times 40$ mesh for a volume fraction, $vf = 0.15$. MaxIter = 1600.}
	\label{fig:Disp_invert_sol_3d}
\end{figure}
The above examples suggest that regardless of the chosen function $f(\beta_j)$ in Eq.~\ref{Eq:nFP_density_beta}, the solutions tend to move towards binary topologies, ultimately producing close-to-binary solutions. Although the function choice does not influence the inclination toward binary topologies, it is noted that the final solution's grayness may vary based on the function selected. One may achieve solutions with gray elements if the function negatively affects gradient magnitudes.

\subsection{Effects of MMA step size}
	This section demonstrates the effects of MMA step size (Eq.~\ref{Eq:MMAapplication}) on the convergence history. We solve inverter problem (Fig.~\ref{fig:Prob_description_disp_invert}), with $f(\beta_j) = 1-\tanh\beta_j$. Additionally, circular neighborhoods with filter radius $r_\text{min} =2$ are employed to illustrate the flexibility of the proposed nFP method in accommodating different neighborhood shapes. The design domain is discretized using $160\times 80$ quadrilateral bi-linear elements. Volume fraction, $v_f = 0.2$ is set. Other parameters are consistent with those in Sec.~\ref{Sec:cg}. We take $S = 0.0025, 0.025,\,\text{and}\, 0.1$ (Eq.~\ref{Eq:MMAapplication}).

The results are presented in Fig.~\ref{fig:inv_solns_circ_DSS}, wherein columns 1, 2, and 3 correspond to $S=0.1$, $S=0.01$ and $S=0.0025$, respectively. The first row displays the convergence plots, while the second row illustrates the optimized mechanisms, with small circles indicating the neighborhood shapes used (Fig.~12). It is observed that for $S=0.1$, the optimization begins to exhibit convergence behavior after 200 MMA iterations for the objective and $g(\bm{\rho})$. In contrast, for $S=0.025$ and $S=0.0025$, convergence starts after 300 MMA and 1600 MMA iterations, respectively. Although the latter two require more computational effort, they result in better-performing mechanisms. Thus, users can find a trade-off between performance and computational cost based on their specific requirements.

\subsection{Extension to 3D-Problems}
Having established the success of the nFP method for different 2D problems, including stiff structures and compliant mechanisms, we extend the method for 3D optimization problems in this section to demonstrate its versatility. The density formulation noted in Sec.~\ref{Sec:density_formulation} is readily modified for 3D problems wherein volumes of the elements consistently replace areas of the elements. The MMA~\cite{svanberg1987method}  is employed for the optimization process herein.

We present two 3D stiff structures (cantilever and MBB beam) and one 3D-compliant mechanism (inverter mechanism) herein. For the presented examples, a neighborhood of an element is defined as the immediate neighbors of the element, that is, elements that share a face, edge, or point with the element. The problem descriptions for the cantilever beam, MBB beam, and displacement inverter mechanism with respective boundary conditions and external forces are shown in Fig.~\ref{fig:Prob_description_3d}. The cantilever beam is symmetric about the vertical mid-plane. The MBB beam is symmetric about the midplane highlighted in gray, while the displacement inverter mechanism problem has 2 planes of symmetry. Given the symmetries, the MBB problem is solved for only half the domain, while the displacement inverter is solved within one-quarter of the domain (highlighted in gray). We do not exploit symmetry conditions for the cantilever beam to demonstrate the method's robustness to provide symmetric results for symmetric problems. The cantilever beam, MBB beam, and displacement inverter problems are solved for a volume fraction of $0.25$, $0.25$, and $0.15$ employing a mesh size of $80 \times 40 \times 40$, $90 \times 30 \times 30$ and $80 \times 40 \times 40$, respectively. An artificial stiffness of $k_a = 100$ is implemented at the output port in the direction of the desired deflection for the displacement inverter problem. 

\begin{figure}[h!]
	\centering
	\subcaptionbox{Cantilever beam: Density distribution. }{\includegraphics[width=0.4\textwidth]{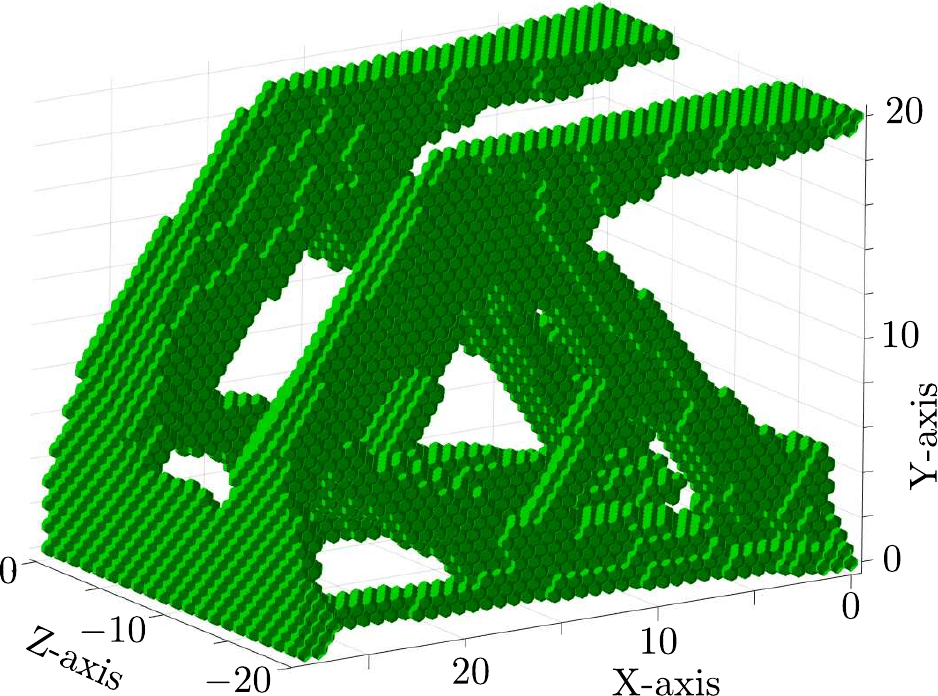}}
	\label{fig:Sol_Canti_nFP_TKD_41x41x41_vf_15_ls_1}
	\hspace{0.5cm}
	\subcaptionbox{Cantilever beam: Convergence history. }{\includegraphics[width=0.4\textwidth]{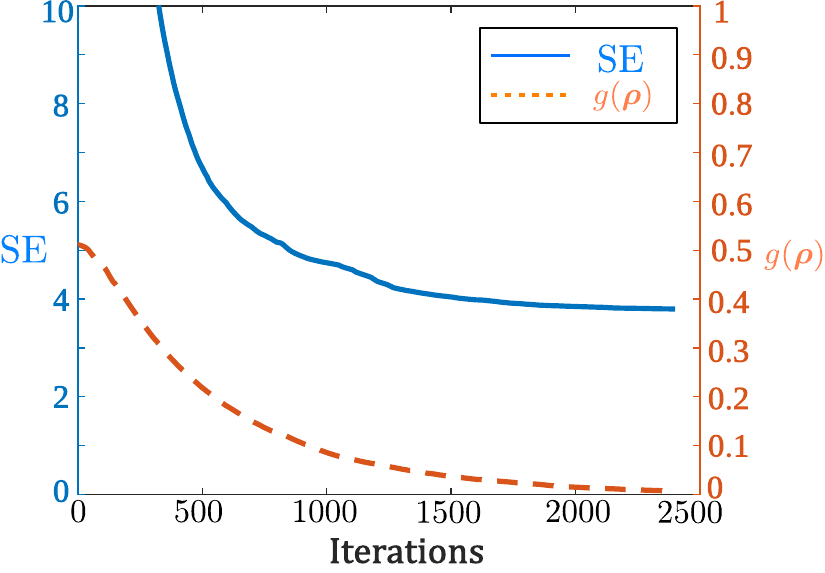}}
	\label{fig:Sol_Canti_nFP_TKD_41x41x41_vf_15_ls_1_convergence}
	\subcaptionbox{Displacement inverter: Density distribution. }{\includegraphics[width=0.4\textwidth]{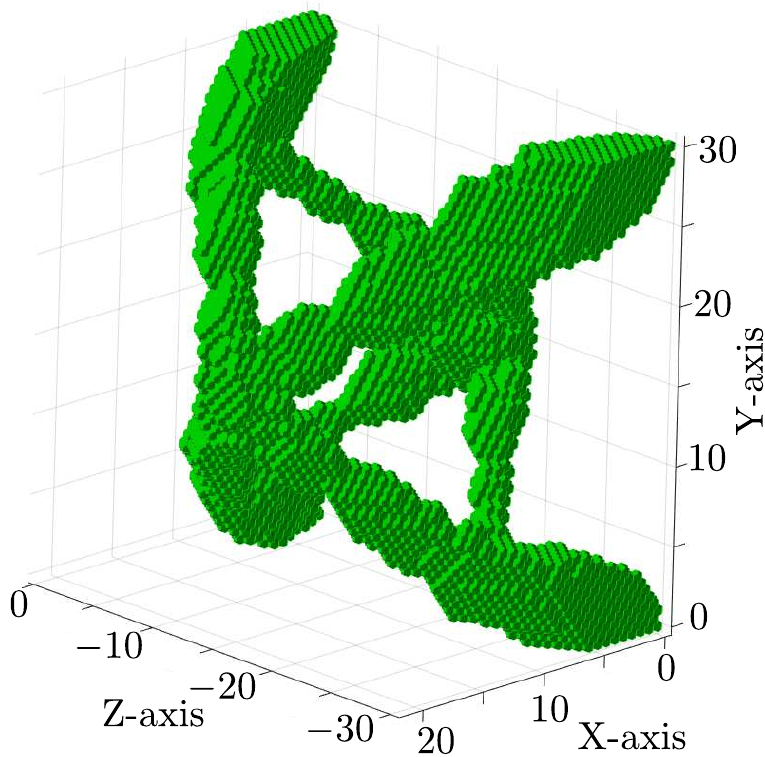}}
	\label{fig:Sol_Disp_invert_nFP_TKD_31x31x31_vf_10_ls_1}
	\hspace{0.5cm}
	\subcaptionbox{Displacement inverter: Convergence history. }{\includegraphics[width=0.45\textwidth]{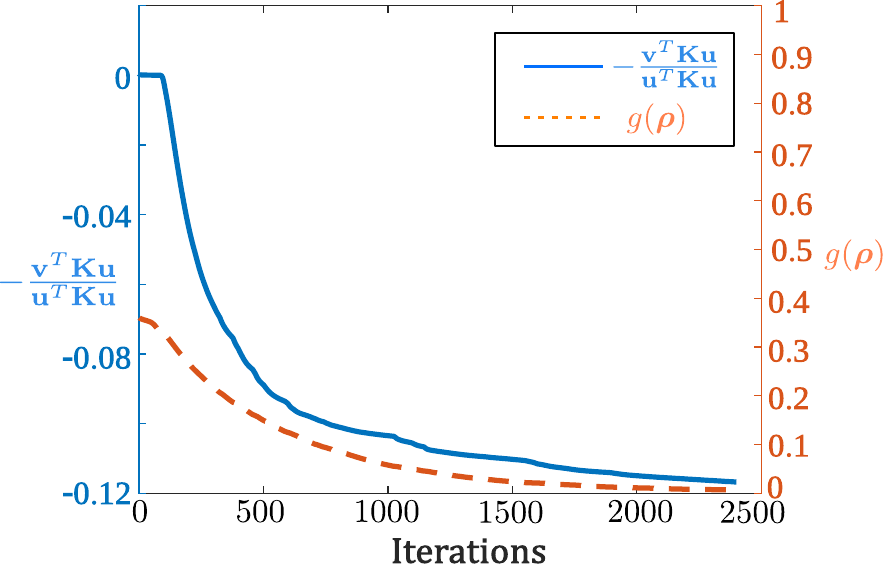}}
	\label{fig:Sol_Disp_invert_nFP_TKD_31x31x31_vf_10_ls_1_convergence}
	\caption{Solutions to the cantilever beam and displacement inverter problems using the nFP approach and truncated octahedron elements. MaxIter = 2400.}
	\label{fig:Sol_nFP_TKD}
\end{figure}

Figure~\ref{fig:Canti_sol_3d}, Fig.~\ref{fig:MBB_sol_3d} and Fig.~\ref{fig:Disp_invert_sol_3d} display solutions and convergence histories for the cantilever beam, MBB beam and displacement inverter mechanism problems, respectively. Though the available symmetry of the cantilever beam is not utilized, the nFP method provides a symmetric optimized cantilever beam displaying its capabilities with respect to symmetry problems. For stiff structures, solutions with element size independent structural members are obtained while local thinning is observed for the displacement inverter mechanism as expected. As noted in 2D problems, for 3D problems, the recorded grayness measures are less than $1\%$, indicating the method's tendency to gravitate towards close-to-binary solutions naturally. 

Like the density method, the proposed nFP method allocates a design variable to each element; thus, with increased mesh refinement, the number of design variables and computational cost increases. We noted that the method can lead to single-point connections in structures. Advanced elements, e.g., hexagonal tessellation in 2-dimensions\cite{saxena2007honeycomb, kumar2022honeytop90} and truncated octahedron tessellations is 3-dimensions \cite{singh2024three} can be implemented to avoid the possibility of singular (point/edge) connections. Next, we show the method's versatility with truncated octahedron tessellations for 3D problems.

\subsubsection{3D results with truncated octahedron elements}
Herein, nFP density evaluation is implemented with truncated octahedron discretization~\cite{singh2024three} to develop a methodology that yields singularity-free, close-to-binary solutions. We solve the cantilever beam and displacement inverter mechanism problems. The considered volume fraction for the former is $0.15$, whereas that for the latter problem is $0.1$. The design domains of the cantilever (full) and inverter mechanism (a quarter) are parameterized using $41 \times 41 \times 41$ and 
$31 \times 31 \times 31$ truncated octahedron elements respectively, with elements of edge length $a=0.25$. The neighborhood of an element contains only the immediate neighbors of the element, that is, elements that share a node with the said element.

The final material layouts, i.e., density distributions, are shown in Fig.~\ref{fig:Sol_nFP_TKD}. The convergence histories are also displayed beside their optimized designs (Fig.~\ref{fig:Sol_nFP_TKD}). The final grayness measures of both solutions are below $1\%$, indicating close to 0-1 solutions. The convergence plots have a similar trend as noted for earlier solutions.
\section{Closing remarks}\label{Sec:Closure}
This paper introduces a novel density evaluation method for topology optimization, relying on the normalized product of a scalar field across a domain. The proposed approach does not require user-defined parameters or weight functions. Instead, users can select a suitable function for their density formulation. The paper outlines the desired properties for such a function and suggests specific choices based on these criteria. The method is demonstrated to generate close-to-binary, transition-free 2D and 3D topologies regardless of the function chosen. The paper also provides this approach's optimization formulation and associated gradient evaluations. 

The density evaluation method, coupled with the SIMP material model, is applied to address compliance minimization and compliant mechanism design problems. The obtained results highlight several key advantages of the proposed density evaluation method: (a) independence from parameter or user choices and continuation scheme, (b) the desired imposition of length scale on a single phase, (c) achievement of mesh-independent solutions, and (d) inherent tendency to generate binary solutions with a suitable function choice. The method's success, efficacy, and versatility are demonstrated on various 2D and 3D stiff structure and inverter mechanism designs for different permitted volumes and length scales. For 3D problems, the method is also demonstrated with advanced octahedral tetrahedral elements. 

We note that the optimization converges faster with respect to the objective; however, the method requires a significant number of iterations to attain a solution with grayness below $1\%$. This opens one of the new avenues to further development. The impact of function choices is also explored, illustrating that the final solution's grayness may depend on this choice for implementation purposes.

Given the numerical experiments performed, the proposed nFP method has performed well with two-phase 2D and 3D problems; extending the method for multi-phase problems can be one of the future directions. The approach can also be extended with design-dependent loading or with finite deformation cases in the near future. 

\section*{Acknowledgments} 
The authors thank  Ole Sigmund for comments  and Krister Svanberg for providing MATLAB codes for the MMA optimizer.
\section* {Conflict of interest} The authors declare that they have no known competing interests.

\section*{Data Availability Statement} The data that support the findings of this study will be made available upon reasonable request.
\appendix
\numberwithin{equation}{section}
\section{Heaviside Projection method and the nFP approach}
\subsection{Gradient comparison}\label{App:A1}
\label{Appendix:A}
Substituting $f(\beta_j) = e^{\beta_j}$ in Eq.~\ref{Eq:nFP_density_beta} yields,
\begin{eqnarray}
	\label{eqn:nFP_den_analytical}
	\rho_i = 1 - \exp \left(\dfrac{\sum\limits_{j \in \mathbb{N}_i}\beta_j A(\Omega_j)}{A(\Gamma_i)}\right)
\end{eqnarray}
where $A(\Gamma_i)$ is area of neighborhood $\Gamma_i$. $\beta_j$ is a \textit{local} parameter, related to element~$j$.  Density formula of the method in Ref.~\cite{Guest2004} is given as,
\begin{eqnarray}
	\label{eqn:Proj_formula_mod}
	\rho_i = 1 - \exp{\left( -\beta \frac{\sum\limits_{j \in \mathbb{N}_i} w_j\mu_j A(\Omega_j)}{\sum\limits_{j \in \mathbb{N}_i} w_j A(\Omega_j)} \right)} + \exp{(-\beta)} \left(\frac{\sum\limits_{j \in \mathbb{N}_i} w_j\mu_j A(\Omega_j)}{\sum\limits_{j \in \mathbb{N}_i} w_j A(\Omega_j)}\right)
\end{eqnarray}
where $\beta \geq 0$ is a user defined \textit{global} parameter, i.e., same for all elements. $w_j$ are the weights obtained using a user defined weight function and $0 \leq \mu_j \leq 1$ are design variables. For a high value of parameter $\beta$, contribution of the third term in Eq. \ref{eqn:Proj_formula_mod} is minuscule. Thus, for a constant weight function and high values of parameter $\beta$ the density evaluation of Ref.~\cite{Guest2004} can be written as:
\begin{eqnarray}
	\label{eqn:Proj_formula_mod_2}
	\rho_i = 1 - \exp{\left( -\beta~ \frac{\sum\limits_{j \in \mathbb{N}_i} \mu_j A(\Omega_j)}{ A(\Gamma_i)} \right)}.
\end{eqnarray}
Density formulae in Eqs. \ref{eqn:nFP_den_analytical} and \ref{eqn:Proj_formula_mod_2} resemble each other. Gradient evaluations for both methods are mentioned  below.
\begin{itemize}
	\item \textbf{Gradient evaluation for nFP}:
	\begin{eqnarray}
		\label{eqn:nFP_grad}
		\dfrac{\partial\rho_i}{\partial \beta_j} = 
		\begin{cases}
			-(1-\rho_i)\dfrac{A(\Omega_j)}{A(\Gamma_i)} &\text{for}~ j \in \{\mathbb{N}_i\}\\
			0 &\text{otherwise}.
		\end{cases}
	\end{eqnarray}
	\item \textbf{Gradient evaluation for Projection}:
	\begin{eqnarray}
		\label{eqn:Proj_grad}
		\frac{\partial \rho_i}{\partial \mu_j} = 
		\begin{cases}
			\beta(1-\rho_i)\dfrac{A(\Omega_j)}{A(\Gamma_i)} &\text{for}~ j \in \{\mathbb{N}_i\}\\
			0 &\text{otherwise}.
		\end{cases}
	\end{eqnarray}
\end{itemize}
For the theoretical case when $\beta \to \infty$, gradients in projection become singular while no singularity is exhibited in nFP when design variables $\beta_j$ approach infinity.
\subsection{Optimized designs comparison}
The mid-load problem (Fig.\ref{fig:Prob_description_MLB}) is selected for this study. Optimized results obtained via the projection method (in elemental form)~\cite{Guest2004} and the proposed nFP for different  $g_{tol}$ are presented below. The publicly available MATLAB code, \texttt{top110}\cite{top110} is employed for the former, wherein the default continuation on $\beta$ is used. We use $f(\beta_j) = e^{\beta_j}$ in density function evaluation for the nFP method.

\begin{figure}[h!]
	\centering
	\subcaptionbox{Iteration 268: $g(\bm{\rho}) = 0.0998$}{\includegraphics[width=0.45\textwidth]{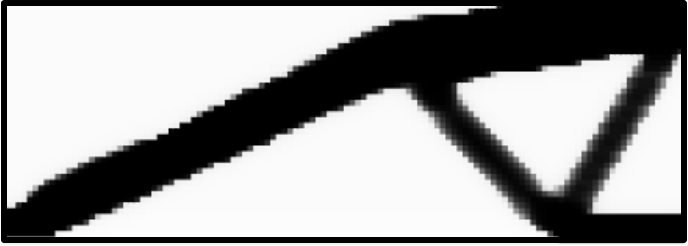}}
	\label{fig:MLB_iter_268}
	\quad
	\subcaptionbox{Iteration 251: $g(\bm{\rho}) = 0.0764$ }{\includegraphics[width=0.45\textwidth]{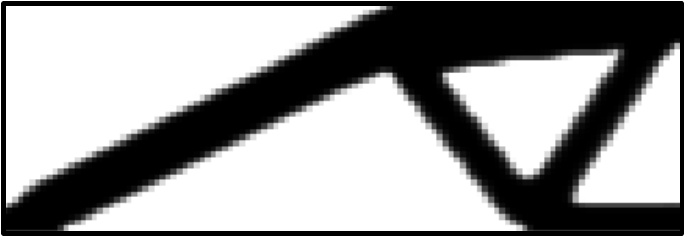}}
	\label{fig:MLB_iter_HDF_251}
	\quad
	\subcaptionbox{Iteration 687: $g(\bm{\rho}) = 0.0710$ }{\includegraphics[width=0.45\textwidth]{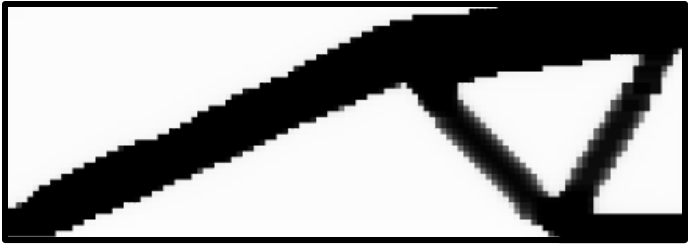}}
	\label{fig:MLB_iter_687}
	\quad
	\subcaptionbox{Iteration 301: $g(\bm{\rho}) = 0.058$ }{\includegraphics[width=0.45\textwidth]{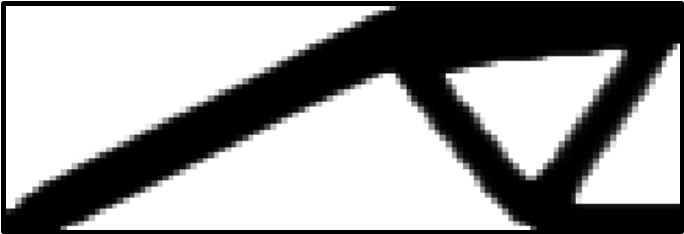}}
	\label{fig:MLB_iter_HDF_301}
	\quad
	\subcaptionbox{Iteration 693: $g(\bm{\rho}) = 0.0472$}{\includegraphics[width=0.45\textwidth]{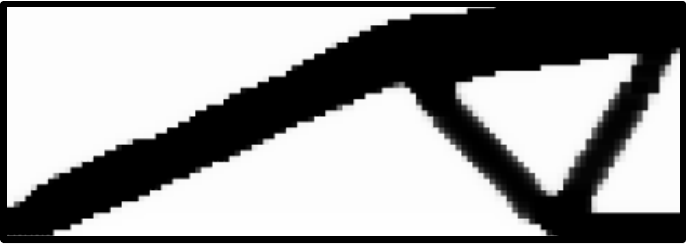}}
	\label{fig:MLB_iter_693}
	\quad
	\subcaptionbox{Iteration 351: $g(\bm{\rho}) = 0.0499$}{\includegraphics[width=0.45\textwidth]{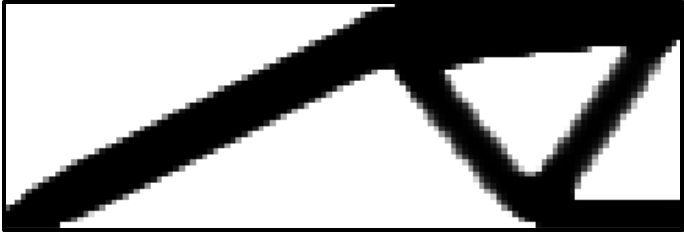}}
	\label{fig:MLB_iter_HDF_351}
	\quad
	\subcaptionbox{Iteration 693: $g(\bm{\rho}) = 0.0208$}{\includegraphics[width=0.45\textwidth]{MLB_693_AP-eps-converted-to}}
	\label{fig:MLB_iter_976}
	\quad
	\subcaptionbox{Does not converge: $g(\bm{\rho})_\text{min} = 0.036$, $\beta_\text{max}= 512$ }{\includegraphics[width=0.45\textwidth]{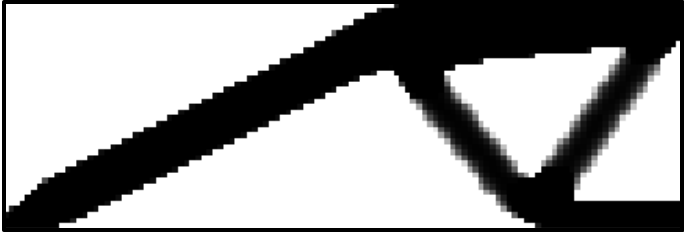}}
	\label{fig:MLB_iter_HDF_400}
	\caption{Results obtained from the proposed nFP method (column~1) and projection method (column 2)~\cite{Guest2004}. Results in row~1, row~2, row~3 and row~4 are obtained with $g_{tol}$ values 0.1, 0.075, 0.050 and 0.025, respectively.}
	\label{fig:Sol_MLB_comparision}
\end{figure}
Column 1 and column 2 of Fig.\ref{fig:Sol_MLB_comparision} provide the results obtained for the mid-load problem using the nFP and the projection method (in elemental form)~\cite{Guest2004}, respectively. The latter method performs relatively better with respect to the required number of iterations for achieving the prescribed $g_{tol}$, as long as $g_{tol}$ is close to 0.05. For $g_{tol} = 0.025$, the nFP converges at $693^\text{th}$ iteration, whereas projection method gets stuck at $g(\bm{\rho})$ = 0.036, even after running over 1000 iterations $g(\bm{\rho})$ does not reduce further. Next, we increase $\beta_\text{max}$ to 2048 and 4096 for the projection method; we find that the optimization process again gets stuck in these cases at $g(\bm{\rho})$ = 0.035 and does not reach $g_{tol}$ value over 1000 iterations. However, such situations are not noted for the nFP approach. The method successfully provides results with  $g_{tol} = 0.025$ (Fig.~\ref{fig:Sol_MLB_comparision}g, Fig.~\ref{fig:Sol_MBB_inter}h),  $g_{tol} = 0.01$ (Fig.~\ref{fig:Sol_MBB_inter}i) and $g_{tol} = 0.0075$ (Fig.~\ref{fig:Sol_MBB_inter}j) at iterations 967$^\text{th}$, 983$^\text{th}$, and 1009$^\text{th}$ iterations, respectively. This numerical experiment confirms that the nFP method will likely reach relatively closer to 0-1 solutions than the projection method~\cite{Guest2004} without using any user intervention/continuation scheme. The cardinal reason may be because the $\beta$ parameter, a global parameter, affects the density variable of all elements in the projection method, whereas, in the nFP method, $\beta_j$ being a local variable focuses on the element to steer the optimization towards close to 0-1. Due to the local nature of $\beta_j$, the nFP method requires a relatively high number of iterations to converge; however, it facilitates solutions relatively more closer to 0-1.
\pagebreak

\end{document}